# Non-Coding RNAs Improve the Predictive Power of Network Medicine

Deisy Morselli Gysi[1,2,3] and Albert-László Barabási[1,2,3,4,*]

[1] Network Science Institute and Department of Physics, Northeastern University, Boston, MA 02115, USA; [2] Channing Division of Network Medicine, Department of Medicine, Brigham and Women's Hospital, Harvard Medical School, Boston, MA 02115, USA; [3] US Department of Veteran Affairs, Boston, MA 02130, USA; [4] Department of Network and Data Science, Central European University, Budapest 1051, Hungary.
* Albert-László Barabási. **Email:** barabasi@gmail.com

**Author Contributions:** A.L.B and D.M.G. designed the study. D.M.G. performed the analysis. All authors read and approved the manuscript.

**Competing Interest Statement:** A.L.B is co-scientific founder of Scipher Medicine, Inc., which applies network medicine strategies to biomarker development and personalized drug selection, and founder of Naring, Inc. which applies data science to health.

**Acknowledgments:** The authors thank Italo F. do Valle and Xiao Gan for fruitful discussions.

**Grants:** This research was supported in part by a National Institutes of Health award 1P01HL132825, a Department of Veterans Affairs award Contract No. 36C24120D0027, and by Scipher Inc. Agreement 21-C-01472.

**Classification:** Biological Sciences: Systems Biology.

## ABSTRACT

Network Medicine has improved the mechanistic understanding of disease, offering quantitative insights into disease mechanisms, comorbidities, and novel diagnostic tools and therapeutic treatments. Yet, most network-based approaches rely on a comprehensive map of protein-protein interactions, ignoring interactions mediated by non-coding RNAs (ncRNAs). Here, we systematically combine experimentally confirmed binding interactions mediated by ncRNA with protein-protein interactions, constructing the first comprehensive network of all physical interactions in the human cell. We find that the inclusion of ncRNA, expands the number of genes in the interactome by 46% and the number of interactions by 107%, significantly enhancing our ability to identify disease modules. Indeed, we find that 132 diseases, lacked a statistically significant disease modules in the protein-based interactome, but have a statistically significant disease module after inclusion of ncRNA-mediated interactions, making these diseases accessible to the tools of network medicine. We show that the inclusion of ncRNAs helps unveil disease-disease relationships that were not detectable before and expands our ability to predict comorbidity patterns between diseases. Taken together, we find that including non-coding interactions improves both the breath and the predictive accuracy of network medicine.

# INTRODUCTION

A key goal of post-genomic medicine is to translate the detailed inventory of cellular components and their disease-related mutations, offered by genomics, into mechanistic insights on disease causation and progression, ultimatelly resulting in novel treatments. To achieve this, we must catalog the physical interactions between all cellular components, arriving at an accurate and predictive map of the human subcellular network. Network Medicine, a discipline whose goal is to exploit the predictive power of subcellular networks[1], has already improved our understanding of disease classification[2] and progression[3], disease-disease comorbidities[4], similarities[4], and treatments[5], and offered tools to identify drug repurposing opportunities[6,7] and drug combinations[8]. Some of these tools have already entered the clinical practice, resulting in network-based diagnostic tool currently used by doctors to improve treatment outcomes for rheumatoid arthritis patients[9], and the drug repurposing opportunities identified during the COVID-19 pandemic[6]. These advances relied on maps of experimentally confirmed protein-protein interactions (PPI), and supported multiple foundational discoveries, like the tendency of proteins associated with the same disease to be co-localized in the same neighborhood of the PPI network[4], the finding that diseases with similar phenotypes lie in similar regions of the interactome[4], and the discovery that the network-based distance of drugs to a disease module affects drug efficacy[5,6].

However, the current interactome maps ignore an important component of subcellular dependency, the regulatory interactions induced and mediated by non-coding RNAs (ncRNAs). In-depth transcriptome sequencing estimates that only 2-3% of the human genome is translated into proteins[10–13], the remaining 97% encoding different classes of non-coding elements[13]. These ncRNAs (Figure 1 A) contribute to multiple biological functions, from the maintenance and regulation of gene expression to pre- and post-processing of mRNAs, splicing, decoding mRNAs into amino acids, and the control of gene expression[14–19], ultimately contributing to multiple diseases[20].

Given the important role these regulatory processes play in disease, a quantitative understanding of disease requires an accurate and comprehensive map of all physical interactions, from the interactions between the proteins, to interactions mediated by non-coding elements (Figure 1 B). The mapping and characterization of the network structure that contains both coding and non-coding elements is necessary to expand the potential of network medicine, leading to better treatments, diagnoses, and ultimately personalized therapies.

Here we respond to this challenge by developing the first comprehensive map of subcellular networks that systematically integrates information on ncRNA mediated interactions with the PPI network, resulting in the first non-coding interactome (NCI). We find that the inclusion of ncRNAs increases the number of nodes by 46% and the number of links by 107% compared to the currently used PPI based interactome. Most important, we find that this expansion allows us to identify disease modules for 132 diseases that lacked a statistically significant module before, hence could not be previously explored with the tools of network medicine. Finally, we show that the expanded interactome improves the prediction of disease-disease relationships, offering more accurate comorbidity predictions, advances that ultimately will lead to better prevention and personalized medicine.

# RESULTS

## PROTEIN-PROTEIN INTERACTION (PPI) NETWORK

The human protein-protein interactome (Figure 1 C) was derived from 21 public databases containing different types of experimentally-derived PPI data[6]: 1) binary PPIs, derived from high-throughput yeast-two hybrid (Y2H) experiments (HI-Union), three-dimensional (3D) protein structures (Interactome3D, Instruct, Insider) or literature curation (PINA, MINT, LitBM17, Interactome3D, Instruct, Insider, BioGrid, HINT, HIPPIE, APID, InWeb); 2) PPIs identified by affinity purification followed by mass spectrometry present in BioPlex2, QUBIC, CoFrac, HINT, HIPPIE, APID, LitBM17, and InWeb; 3) kinase-substrate interactions from KinomeNetworkX, and PhosphoSitePlus; 4) signaling interactions from SignaLink and InnateDB; and 5) regulatory interactions derived by the ENCODE consortium. We used the curated list of PSI-MI IDs provided by Alonso-López *et al* (2019) to differentiate binary interactions among the several experimental methods present in the literature-curation databases. Specifically, for InWeb, interactions with curation score < 0.175 (75$^{th}$ percentile) were not considered. All proteins were mapped to their corresponding Gene Symbol (NCBI) and proteins that could not be mapped were removed. As each database contributes with interactions between a different set of proteins (Table S1), the resulting PPI network contains 536,965 interactions between 18,217 proteins (Figure 1 E). Interactions containing at least one ncRNA transcript were included in the non-coding interactome (NCI).

**NON-CODING INTERACTOME (NCI)**

The most studied ncRNAs are microRNAs (miRNAs), that contain ~22 nucleotides[18] and mainly act at the post-transcriptional level[19], involved in mRNA cleavage, activating or repressing translation[17,18]. The human genome accounts for approximately 2,300 miRNAs[21], each with hundreds of targets, which together regulate 10 to 30% of all genes[22]. miRNA recognizes its mRNA targets by base-pairing the miRNA seed region (containing 2-8 nucleotides) to the complementary region on the targeted mRNA[23] (Figure 1 B). Playing a similar role as Transcription Factors (TFs), miRNAs form an intertwined network[24] that affect disease development[25] as documented in asthma[26] or schizophrenia[27], and mutated or dysregulated miRNAs are associated with the lack of function in neurogenesis[28].

Long non-coding RNAs (lncRNA), another family of nonprotein-coding RNAs, that exceed 200 nucleotides[29], present a poli-A tail, and can be spliced. Even though only a small number of lncRNAs are well-characterized[30], they are involved in a wide range of biological functions, from X-chromosome inactivation[31,32], to imprinting[33,34], and often interact with proteins[35–37] – acting as a TF, and sponges for miRNA[38], bind to chromatin[39] and enhancers[40]. lncRNAs can bind to both RNAs and proteins, hence are classified into protein-focused and RNA-focused[30] (Figure 1 B). Moreover, lncRNAs are also associated with multiple diseases[41], cancers[42], autoimmune neuropathies[24], and neurodegenerative diseases[43], and the lncRNA CRNDE has been identified as a promising target for the therapeutic treatment of prostate cancer[44].

To construct the human non-coding interactome (NCI, Figure 1 D) we combine nine publicly available databases that collect and curate experimentally-derived non-coding interactions: 1) miRNA-targets, derived from reporter assay, western blot, microarray, and next-generation sequencing experiments from mirTARbase; 2) miRNAs and lncRNAs interactions validated using CLIP-seq, AGO CLIP-seq, ChIRP-seq, HITS-CLIP and PAR-CLIP from lncBook, NPinter4, DIANA Tools, RAIN, and lncRNome; 3) RNA-RNA interactions validated from transcriptome-wide sequencing-based experiments (PARIS, SPLASH, LIGRseq, and MARIO), and targeted studies (RIAseq, RAP-RNA, and CLASH) from RISE; 4) Literature curated from miRNet and miRecords. Additionally, we include any PPI-derived interaction involving at least one ncRNA. A detailed description of the databases can be found in SI 2.1.

Given our focus on experimentally validated interactions between protein-coding or non-coding genes, we did not include databases limited to predicted or literature-mined interactions without experimental validation, such as mirwalk, TargetScan, PicTar, TargetMiner, TargetScanVert, miRDB, and microrna.org. Finally, several other classes of ncRNAs can help maintain the homeostasis of the cell. For example, small nuclear RNAs (snRNAs) help pre-process mRNA, performing splicing, or intron removal; Transfer RNAs (tRNAs) help decode mRNAs into peptides or proteins; Ribosomal RNAs (rRNAs) are involved in protein translation; housekeeping RNAs, such as rRNAs, can carry modifications (e.g. methylations and pseudouridylations), guided by small nucleolar RNAs (snoRNAs)[16], and small double-stranded RNAs (dsRNAs) mediate post-transcriptional gene silencing of mRNAs, via RNA interference (RNAi). Currently we lack databases that curate their interactions with other cellular components, limiting our ability to systematically explore their role.

**NETWORK ANALYSIS**

We begin by constructing two networks: i) the protein-protein interaction (PPI network), which contains 536,965 interactions between 18,217 protein-coding genes, and ii) the joint PPI & NCI network, which has 26,575 genes (18,358 coding, 2,134 lncRNA, 1,650 miRNAs, 3,429 pseudogenes and 1004 other ncRNAs such as piRNA, siRNA, tRNAs) connected by 1,114,777 binding interactions (Figure 1 E). The inclusion of ncRNAs increases the number of nodes by 46% and the number of links by 107%, compared to the PPI. The final interactome is fairly complete, containing 94,5% of all human proteins, 99.6% of all TFs, 86,3% of all miRNAs, and 38.5% of all lncRNAs transcripts (Figure S4).

While in the PPI proteins interact with 30 [12; 64] (median value [interquartile range]) other proteins, in the PPI & NCI network the median degree increases to 54 [20; 107]. A TF in the PPI interacts with 48 proteins [20; 102] and its degree increases to 90 [41; 168] after the inclusion of the ncRNAs. miRNAs and lncRNAs, absent in the PPI, are connected to 52 [16; 224.75] and 3 [1; 6] elements (miRNAs, lncRNAs, proteins, TFs, or other ncRNAs) respectively (Table S2). In the PPI & NCI network, 34.8% of the interactions are between miRNAs and protein-coding genes and 16% between miRNAs and TFs, and lncRNAs are responsible for less than 1% of the combined interactions (Figure 1 E). In summary, we find that miRNAs are responsible for the majority of the non-coding interactions with protein-coding genes, TFs, and pseudogenes, as well as 40% of the lncRNAs interactions (Figure 1 E), confirming the central role miRNAs play in the regulation of the underlying cellular network.

**DISEASE-DISEASE ASSOCIATIONS UNDER THE LIGHT OF NON-CODING RNAS**

Relating genes and their mutations to diseases is the central question of post-genomic medicine, drug-based therapeutics, drug discovery, and drug repurposing. Genes are typically associated with traits using genome-wide association studies (GWAS) combined with functional analysis to evaluate the effect of a single nucleotide polymorphism (SNP) or gene on the trait. We assembled a Gene Disease Association (GDA) database by retrieving evidence of strong disease associations (GWAS followed by experimental validation) or weak disease associations (GWAS from ClinGen, ClinVar, CTD, Disease Enhancer, DisGeNET, GWAS Catalog, HMDD[45], lncBook, LncRNA disease, LOVD, Monarch, OMIM, Orphanet, PheGenI, and PsyGeNet, see Material and Methods, SI Section 2.2, and Table S3).

We limit our exploration to diseases with at least 10 strong or weak gene associations, arriving to a database with 861 diseases and 13,216 disease-associated genes, of which 10,764 are protein-coding and 2,452 encode ncRNAs. The diseases with most miRNAs associations are carcinoma hepatocellular (380 miRNAs), breast neoplasms (372 miRNAs), and colorectal neoplasms (347 miRNAs). Diseases with high lncRNAs involvement include astrocytoma (311 lncRNAs), breast neoplasms (114 lncRNAs), and stomach neoplasms (112 lncRNAs). We find that 250 of the 861 diseases are enriched for ncRNAs (proportions test; p-adj < 0.05, FDR corrected), i.e., they have more associated ncRNAs than we would expect by chance.

Previous results show that proteins associated with a specific trait, phenotype, disease, or biological process tend to cluster in the same topological neighborhood of the PPI network[46], forming a sub-network known as the disease module. The size of the largest connected component (LCC) of this sub-network measures the magnitude of the disease module, and its statistical significance is obtained by deriving the LCCs distribution based on the random selection of the disease-associated genes. As a second measure, here we introduce the relative LCC (rLCC), defined as the ratio between the size of LCC and the number of genes associated with the disease. The rLCC captures the completeness of the disease module, allowing comparison across diseases with different numbers of genes. For example, schizophrenia has 1,458 associated genes (1,292 coding, 148 ncRNAs) of which 1,364 form an LCC in the PPI & NCI ($p < 0.05$), resulting in an rLCC of 93,55%. In contrast, even with the inclusion of ncRNAs, obsessive-compulsive disorder, with 92 genes (81 protein-coding and 11 non-coding) has an LCC of 8 in the PPI & NCI, resulting in an $rLCC_{Ppi \& NCI}$ of only 8.69%, indicating that the corresponding disease module is highly incomplete.

In the following we focus on three diseases, Rheumatoid Arthritis (RA), Chron's Disease (CD), and Pre-Eclampsia (PE), to illustrate the value of adding the NCI to the interactome. RA is a multisystemic, chronic inflammatory disease characterized by destructive synovitis, erosive arthritis, progressive articular damage, and systemic organ involvement[47–49], which can lead to decreased functional capacity and quality of life, increased morbidity and mortality. RA is associated with 391 genes, (302 protein-coding, 66 miRNAs, 14 lncRNAs, and 9 other ncRNAs). While there is a significant disease module (p-adj < 0.05, FDR corrected) in both the PPI and the PPI & NCI, the PPI's disease module accounts for only 189 disease-associated genes ($rLCC_{PPI}$: 62%), whereas the PPI & NCI includes 345 disease-associated genes (Figure 2 A), increasing the $rLCC_{PPI\ \&\ NCI}$ to 88%.

Chron's Disease (CD) is an autoimmune disease, that causes inflammation of the digestive tract, which can lead to abdominal pain, severe diarrhea, fatigue, weight loss, and malnutrition. CD is associated with 279 genes (204 protein-coding, 47 miRNAs, and 14 lncRNAs). It has a significant disease module in both networks, however, the size of the LCC more than doubles with the inclusion of the ncRNA ($LCC_{PPI}$: 93; $LCC_{PPI\ \&\ NCI}$: 217), and the proportion of disease-associated genes also increase greatly (from $rLCC_{PPI}$: 45% to $rLCC_{PPI\ \&\ NCI}$: 77%) (Figure 2 B), collecting many more known disease genes.

Finally, pre-eclampsia (PE) is characterized by persistent high blood pressure during pregnancy or postpartum, and in rare cases, it can progress to severe PE very quickly, which can lead to the death of the mother and the baby. PE can also lead to premature birth, malnutrition, and lack of oxygen in the womb; adults whose mothers had PE have higher chances of developing diabetes, congestive heart failure, and hypertension[50]. PE has no statistically significant disease module in the PPI network, hence, previously we could not apply network medicine tools to explore this disease. Indeed, PE is associated with 136 genes; the majority of which (95) encode miRNAs. It does, however, have a significant module in the combined PPI & NCI network (Figure 2

C). The PPI network accounts for only 33 of the 136 genes associated with the disease, of which only 2 are part of the LCC, while the PPI & NCI contains 136 genes of which 111 are in the LCC. Similarly, the ratio of disease genes found in the disease module jumps from 6% ($rLCC_{PPI}$) to 81% ($rLCC_{PPI\ \&\ NCI}$), confirming the key role miRNAs play in regulation of PE. In other words, the PPI, historically the starting point of network medicine studies, provides a highly incomplete map of the PE disease module. However, the inclusion of ncRNAs allows us to now detect the disease module, opening up the possibility to explore the disease using the tools of network medicine.

To generalize beyond RA, CD, and PE, we calculated the LCC, rLCC, and the significance of the disease module for all 861 diseases in the PPI and in the PPI & NCI. Of the 861 diseases, 505 have a statistically significant LCC in the PPI network and 602 have a statistically significant LCC in the PPI & NCI (permutation test, N = 1000, p-adj < 0.05, FDR corrected; Figure 3 A). Taken together, we find that for 132 diseases the identification of a statistically significant disease module cannot be achieved without the inclusion of ncRNA-based interactions. We also find that the inclusion of ncRNAs decreases the LCC p-value (FDR corrected) for 522 diseases, hence increasing our confidence over the observed disease module (Figure 3 B, C, E), and also increases the median size of the significant LCCs (Wilcoxon Test, p < 0.0001; Figure 3 D). Finally, we find that the rLCC in the PPI & NCI increases for 430 diseases by 26.6% on average, indicating that the NCI considerably reduces the number of unconnected disease genes by linking many previously isolated components to the LCC.

Taken together, we find that including ncRNAs in the PPI increase the size, significance, and disease-gene retrieval of the disease module, expanding the reach of network medicine to a large number of diseases that previously could not be explored using network-based tools. Digestive problems, pathologies, and neoplasm have the highest ratio of ncRNAs (Figure 3 F), indicating that those disease classes benefit most from the inclusion of ncRNAs.

**DIRECT AND INDIRECT BINDINGS ARE SUPPORTED BY CO-EXPRESSION NETWORKS**

Gene co-expression networks are often used to shed light on the molecular mechanisms that underlie biological processes[51]. As gene co-expression is driven by the regulatory network, it leads to the hypothesis, supported by previous evidence[52], that interacting proteins are products of genes with higher co-expression, compared to proteins that do not physically interact. Here we use gene co-expression to probe the relative strength of the interactions induced by coding and non-coding elements. As bulk RNA-seq is not appropriate for measuring miRNA, due to the lack of poli-A tail in miRNAs, here we derive an indirect physical network that includes co-regulatory interactions (Figure 4 A), driven by the hypothesis that two proteins are co-expressed if they are co-regulated by the same ncRNAs. We distinguish three interactions that could modulate co-expression networks: i) Direct PPI (Figure 4 B), constructed using direct physical interaction between two proteins; ii) Indirect NCI (iNCI; Figure 4 C), connecting two proteins if proteins "A" and "B" are co-regulated by the same ncRNA; iii) Direct & iNCI (Figure 4 D), representing the combination of the PPI and the iNCI network; or iv)

For control, we measure co-expression between nodes that have no known physical interaction between them.

We compare the measured co-expression for the three different interaction types (PPI, iNCI, and PPI & iNCI), relying on gene co-expression derived from whole blood samples by GTEx[53]. We use both Pearson Correlation ($\rho$), and wTO[54,55] ($\omega$) to ensure that the results are not biased by the methodology. We then compare the absolute co-expression weights ($\rho$ and $\omega$) for the three binding interaction types (PPI, iNCI, and PPI & iNCI, and no interaction). We find that the absolute co-expression weights ($\rho$ and $\omega$) are, on median, higher for all three types of physical interactions (Figure 4 E, Figure S6 A) compared to the control (Kruskal-Wallis; Dunn's posthoc test, p-adj < 0.05 Holm method; Table S4), confirming that two protein-coding genes that interact (directly or indirectly) have higher co-expression compared to genes that do not interact. Most important, we find the strength of the co-expression induced by the PPI or by the iNCI to be statistically indistinguishable, indicating the comparable impact of non-coding interactions on coexpression.

The observed higher co-expression values on the physical networks (PPI, iNCI, and PPI & iNCI), prompt us to ask whether the co-expression weights are predictive of physical binding. We, therefore, calculated the Area Under the ROC (AUROC), which measures the ability to discriminate between binding or no binding, where an AUROC of 0.5 indicates a random choice (i.e., lack of predictive power), while an AUROC of 1 indicates accurate predictions. We find that the AUCs increased from 0.59 in the PPI to 0.63 in the iNCI and PPI & iNCI networks for $\rho$ (Figure 4 F) and from 0.58 PPI to 0.63 in the iNCI and PPI & iNCI networks for $\omega$ (Figure S6 B). In other words, the inclusion of ncRNA-induced indirect interactions improves the accuracy of correlation-based networks to predict physical interactions, demonstrating the important role NCI play in the interpretability of co-expression patterns.

**DISEASE COMORBIDITY AND DISEASE PROGRESSION**

Diseases with similar phenotypes tend to have common genetic roots, as captured by the Jaccard Index of genes associated with different phenotypes. In addition, diseases with similar symptoms tend to share their disease network neighborhood[56], a feature captured by the network-based separation of two diseases, $a$ and $b$, defined as[4]

$$S_{a.b} = <d_{a,b}> - \frac{<d_{a,a}> + <d_{b,b}>}{2},$$

where $<d_{i,j}>$ is the average shortest distance from disease $i$ to $j$. A negative $S_{a.b}$ indicates that two diseases are in overlapping network neighborhoods, while $S_{a.b} \geq 0$ implies that components associated with diseases $a$ and $b$ are in distinct network neighborhoods.

To illustrate how NCI can help improve our understanding of disease relationships, we focus first on RA and its comorbidities. If we limit the network to the PPI, RA has only one statistically overlapping disease, Chron's Disease (Figure 5 A and B – PPI). In contrast, in the extended PPI & NCI network, $S_{a,b}$ uncovers statistically significant network-based overlap with Chron's Disease, Ulcerative Colitis, Inflammatory Bowel Disease, inflammation, Systemic lupus erythematosus (SLE), multiple sclerosis (MS), diabetes mellitus types 1 and 2, asthma, heart failure, stroke, atherosclerosis and multiple neoplasms (Figure 5 A and B – PPI & NCI). Most of these diseases predicted to be in the same network neighborhood as RA have known comorbidities to RA, confirmed by clinical evidence. Indeed, RA patients present chronic inflammation[57], often develop SLE, a combination known as Rhupus[58,59]; MS is a well-known comorbidity of RA[60], similarly to inflammatory bowel diseases[61] (e.g., Ulcerative Colitis and Chron's Disease[62]). Also, patients with RA have a higher risk of developing diabetes type I[63] due to insulin resistance, and patients with RA often develop diabetes mellitus[64], additionally, patients with asthma have higher risk of developing RA[65,66]. Moreover, patients with RA have reported higher rates of heart failure[67,68], such as myocardial infarction[67] and stroke[68] in addition to atherosclerosis[69]. Higher risk of multiple neoplasms have also been reported in patients with RA treated with anti-TNF drugs[70,71], interestingly, even though multiple myeloma and colonic neoplasms[72] share the same network neighborhood, they have reported decreased risk in patients with RA[73], suggesting that two diseases in the same network vicinity might also grant protection from each other. Taken together, we find that while the PPI can identify only one comorbidity for RA, most of the clinically documented comorbidities can be detected in the joint PPI & NCI network, indicating that the inclusion of ncRNA interactions is necessary to reveal disease comorbidities.

As a second case study, we focus on pre-eclampsia (PE), which lacks a disease module in the PPI, hence, we could not predict comorbidities based on the PPI alone. By using the PPI & NCI combination, the top 10 closest diseases with significant topological overlap with PE are atherosclerosis, uterine cervical neoplasms, osteosarcoma, pancreatic neoplasms, glioma, cholangiocarcinoma, multiple myeloma, triple-negative breast neoplasms, heart failure and carcinoma pancreatic ductal. Indeed, there is clinical evidence that women with PE are at increased risk for atherosclerosis[74] and other cardiovascular diseases such as heart failure[75], women with HPV infection, the main cause for uterine cervical neoplasms, also have increased risks for PE[76], and PE has been associated to an increased risk for several types of cancer[77]. In other words, the joint PPI & NCI can accurately predict the known comorbidities of PE.

Finally, we expand our investigation by mapping out the disease-disease relationships that capture the network proximity of all disease-pairs. The PPI predicts 543 comorbidity links between 350 diseases, revealing distinct clusters for neoplasm, cardiovascular and gastrointestinal disease (Figure 5 C – PPI). In the combined PPI & NCI, we find 2,659 pairwise disease links between 466 diseases (Figure 5 C – PPI & NCI). We find that by including ncRNAs into the PPI we retrieve more complete and biologically more meaningful list of disease-disease interactions, offering a better quantitative understanding of disease similarity and comorbidity, ultimately helping us understand disease progression in patients[3].

**THE NON-CODING INTERACTOME PREDICTS RELATIVE RISK**

The PPI-based disease-disease separation is a known predictor for disease comorbidity[4], prompting us to ask whether the inclusion of ncRNAs improves not only our ability to detect clinically documented comorbidity in patients but can also help quantify comorbidity by predicting the relative risk between diseases. We measured the pairwise relative risk (RR) between all disease-pairs, relying on the disease history extracted from 13,039,018 elderly Americans enrolled in Medicare[78]. RR estimates the strength of the association between two diseases, so that RR > 1 represents a risk factor, while RR < 1 indicates a protective factor. For example, Ulcerative Colitis (UC) affects 26,432 patients and Crohn's disease (CD), 24,936. If the two diseases were independent of each other, we would expect only 45 individuals with both diseases. In contrast, we find 1,462 individuals with both UC and CD in our database. In other words, the chance of a patient developing CD is 30.55 times higher [29.0, 32.16; Wald Interval, 95% confidence] in a patient with UC compared to a patient that does not have a history of UC, meaning that UC is a risk factor for CD.

Next, we investigate if diseases with a network overlap ($S_{a,b} < 0$) have a higher RR, meaning that diseases that are located in the same network neighborhood have a higher risk of being comorbid. We find that negative $S_{a,b}$ disease pairs have a statistically higher RR than the ones with positive $S_{a,b}$ in both the PPI and the PPI & NCI (see Methods, Figure S7 A). The RR in the PPI is on average 8.6 (se 2.98) for diseases with an $S_{a,b} < 0$ (p < 0.05), and 6.74 (se 0.41) for diseases with an $S_{a,b} > 0$ (p < 0.05). For the PPI & NCI, we observe an increase in the RR average to 9.5 for negative $S_{a,b}$ (se 2.93), and a decrease to 6.5 (se 0.41) for positive $S_{a,b}$ (Figure S7 A). The increase of the RR for closer diseases in the PPI & NCI networks indicates, once again, that the inclusion of ncRNAs enhances our ability to quantify comorbidity by predicting the relative risk for patients, offering a better understanding of the network-based roots of disease progression.

## DISCUSSION

Network medicine, with its promise to offer a better mechanistic understanding of diseases[4], their progression[3], comorbidities[4], similarities[4], and treatments, such as drug repurposing[6,7] and drug combinations[8], traditionally relied on protein-protein interactions (PPI), capturing binding interactions between proteins. Two decades after the Human Genome Project, there is overwhelming evidence that non-coding genes and ncRNAs regulate multiple biological processes and functions, playing important roles in multiple diseases, hence must be incorporated into the network medicine framework.

Here, we show that the inclusion of ncRNAs into the PPI significantly improves the breath and the predictive power of network medicine. Protein-coding and non-coding genes are intertwined into a densely connected network, hence the inclusion of ncRNAs improves disease module identification and our ability to uncover disease-disease relationships, more accurately predicting the relative risk for patients. We also show that the rLCC increases when we incorporate the ncRNAs, helping us retrieve more complete disease modules, and offering biologically more interpretable identification of the molecular components contributing to a disease. We find that several neoplasms share the same non-coding neighborhoods, confirming the role ncRNAs play in neoplasm regulation. We also find improved comorbidities for many other diseases after the inclusion of ncRNAs, suggesting that non-coding elements contribute to most disease mechanisms.

The more accurate disease module detection enabled by the inclusion of ncRNAs can lead to the development and identification of novel drug-targets, that hit closer to the disease module. They also raise the possibility that for some diseases targeting ncRNAs in the disease module may have a better therapeutic potential than targeting proteins. The clinical relevance of such intervention is illustrated by Bevasiranib[79], a small interfering RNA (siRNA) that targets the VEGF-A gene, currently in clinical trial for treating macular degeneration and diabetic retinopathy, or Inclisiran, an LDL cholesterol lowering siRNA that targets PCSK9, the first approved ncRNA-based drug[80,81]. We also found an improved comorbidity prediction when considering ncRNAs, suggesting that the systematic inclusion of non-coding elements can offer a better understanding of disease progression, potentially opening a path towards precision medicine.

# MATERIALS AND METHODS

## GENE DISEASE ASSOCIATIONS

We surveyed around 130 databases with Gene-Disease-Associations (GDA) and selected those that i) were not compiled from other data sources, and ii) provided at least one kind of evidence type classified as: Strong (functional evidence using an experimental essay); Weak (GWAS evidence but no experimental validation); Inferred (relying on bioinformatics or SNPs from imputation in GWAS); not compatible ((l)ncRNA, miRNA and other transcripts with or without experimental validation). For each database we kept the disease name, gene converted to HGNC names (HUGO Gene Nomenclature Committee), and evidence level. At the end, we combined the following data sources: GWAS from ClinGen, ClinVar, CTD, Disease Enhancer, DisGeNET, GWAS Catalog, HMDD[45], lncBook, LncRNA disease, LOVD, Monarch, OMIM, Orphanet, PheGenI, and PsyGeNet (See SI 2.2).

We searched for datasets that provide non-coding interactions derived from experimental evidence (ncRNA vs proteins or ncRNA vs ncRNA). We kept only databases that provided experimental evidence for their binding interactions and removed all interactions without strong evidence (see SI 2). To validate the collected interactions, we assessed each database's interactions and gene's overlap.

**COMBINING DATA**

All disease names were converted into MeSH terms after a word2vec embedding, and all gene IDs into Gene Symbols. Gene names were normalized to HGNC symbol using biomaRt[82], Gene Cards[83], and gene2ensembl from NCBI. Genes were classified into coding, and non-coding (miRNA, ncRNA, etc) based on their classification from Gene Cards. Coding genes were classified as Transcription Factor (TF) according to Perdomo-Sabogal 2019[84].

To normalize disease names, we first converted all strings to low-case and kept only alphanumeric characters. We next removed diseases with keywords that represented measures (such as "body mass", "volume", "count", "susceptibility", etc). The renamed diseases were combined into clusters based on their similarity to MeSH C or F terms and synonyms based on a word2vec embedding trained on PUBMED[85]. Disease names with a cosine distance lower than 0.8 to a MeSH term or a MeSH synonym were removed, and the term with higher similarity was selected. A disease classification is based on the MeSH's first level of classification. After disease and gene names normalization, we filtered for diseases that have at least five associated genes, with at least one strong, weak, or incompatible experimental evidence.

**NETWORK MEDICINE TOOLS**

Disease Modules were inferred from the gene-disease association curated database, by calculating the Largest Connected Component (LCC) of each disease, and deriving its p-value from the density of 1000 simulations of a permutation test. LCCs were estimated using the NetSci R package.

Diseases separation was calculated using the measure proposed by Menche et al 2015[4], and implemented in the R package NetSci. Separation significance is calculated by resampling 1000 times the genes in each disease, and calculating the $S_{a,b}$, one-sided p-values are calculated as the density from $-\infty$ to the found $S_{a,b}$.

To assess the disease similarity network, we selected overlapping elements given their disease separation[4] ($S_{a,b}$) < 0, significance of the $S_{a,b}$ (permutation test, N = 1000; p < 0.05), significance of the gene overlap (Hypergeometric test, p-adj < 0.05; FDR corrected) and Jaccard Index > 0 for each disease pair that forms a significant LCC using the two pre-defined networks (PPI, and PPI & NCI).

**DRUG-TARGETS AND GENE CO-EXPRESSION**

Drug-Target interactions were retrieved from DrugBank[86] (version 5.1.9), keeping all target types (polypeptides, enzymes, carriers, and transporters). We filtered for drugs that have at least one drug-target described.

Gene co-expression networks were inferred using the association between pairwise gene expression, measured using RNAseq or microarray, and the association is often derived from a correlation, such as Pearson correlation, or a transformation, such as the Weighted Topological Overlap (wTO)[54,55]. Gene expression was accessed using whole blood samples from GTEx[53], and the co-expression was constructed using

Pearson Correlation and the Weighted Topological Overlap, from the wTO R Package[54], which calculates the co-expression between two genes based on their normalized correlation and further removes false positives.

**DIRECT AND INDIRECT PHYSICAL INTERACTIONS**

We created an indirect-physical interaction network, which identifies the co-regulation of any two protein-coding genes by the same non-coding gene. For that, we constructed a bipartite network, based on the PPI & NCI, where one set of nodes represented non-coding genes and the other set represented the protein-coding genes. From this network, we created the projection on the protein-coding network, identifying if two protein-coding genes are targets of the same ncRNAs. We combine the original PPI with the indirect-physical interaction network.

**STATISTICAL ANALYSIS AND RELATIVE RISK**

Statistical analyses were performed using the R environment. Tests and confidence intervals have been reported along with their p-values. All tests, unless stated, were FDR-corrected.

To access the relative risk of two diseases (comorbidities) we used the data from patients enrolled in MediCare[4,78]. The complete network using 3 ICD digits was first converted into MeSH codes using the procedure described in the Disease Association. MeSH codes could be matched to multiple ICD codes, and therefore patients were combined into the identified MeSH category. Relative Risk, and confidence interval were accessed using the RelRisk function from the DescTools R package, using the Wald Test.

To access the difference between the RR and the $S_{a,b}$, we select RR > 1 with statistical significance (p < 0.05, FDR corrected), and focus on diseases that affect at least 5% of the population, avoiding inflated RR. Differences between groups were accessed using the Mann-Whitney test.

# FIGURES

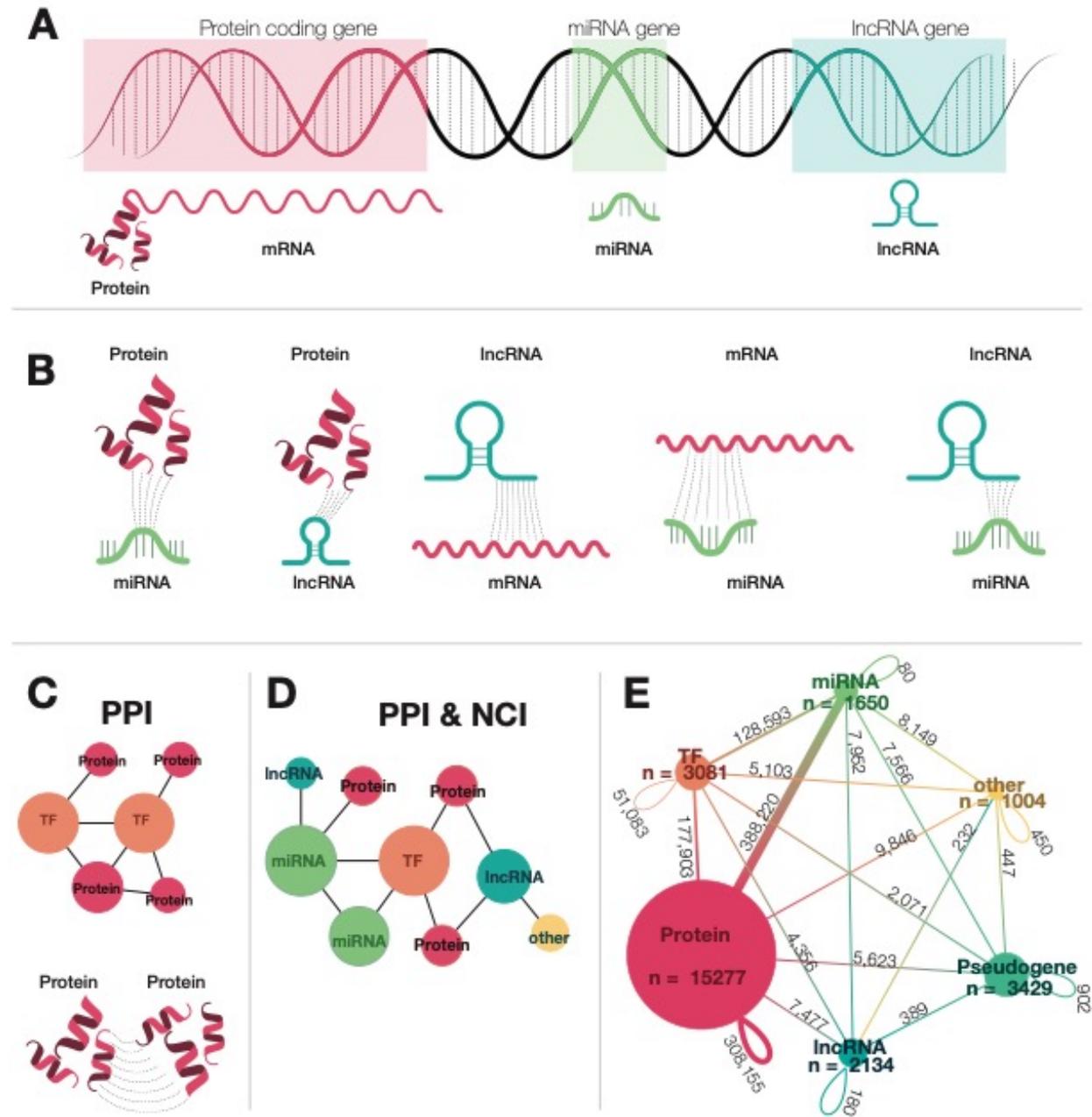

Figure 1 **The role of ncRNA in gene regulation and connection to the human interactome. A)** The Modern Central Dogma of Biology. A DNA strand showing the transcription process: miRNAs, lncRNAs, and mRNAs are all transcribed from the DNA, however, only processed mRNAs are translated into proteins. **B)** Interaction Between ncRNAs. miRNAs can bind to lncRNAs, mRNAs, and proteins. When miRNAs interact with mRNAs and lncRNAs they regulate (by activating or repressing) the gene expression process. lncRNAs can also bind to miRNAs, mRNAs, and proteins. **C)** A Network of Protein Interactions. Proteins interact with one another, forming a Protein-Protein Interaction network. Some proteins act as transcription factors (TF), which regulate gene expression. The PPI only accounts for binding

interaction among protein-coding genes. **D)** A Network of All Interactions. ncRNAs and protein-coding RNAs interact with each other, forming a densely connected network, the PPI & NCI, which contains multiple types of physical interactions from different genomic elements. **E)** PPI & NCI. Each edge on the network represents the relative frequency of all respective interactions across different element types. The PPI is a subgraph of the PPI & NCI, which only accounts for protein-coding genes and their interactions; TFs and Proteins interact with each other, responsible for 33% of the interactions in the PPI & NCI network, showing that even though protein interactions play a big role on the network, their interaction with other groups is also important. The majority of interactions occur between miRNAs and protein-coding genes and TFs. While lncRNAs interact with protein-coding genes and other transcription factors, they interact with few other elements.

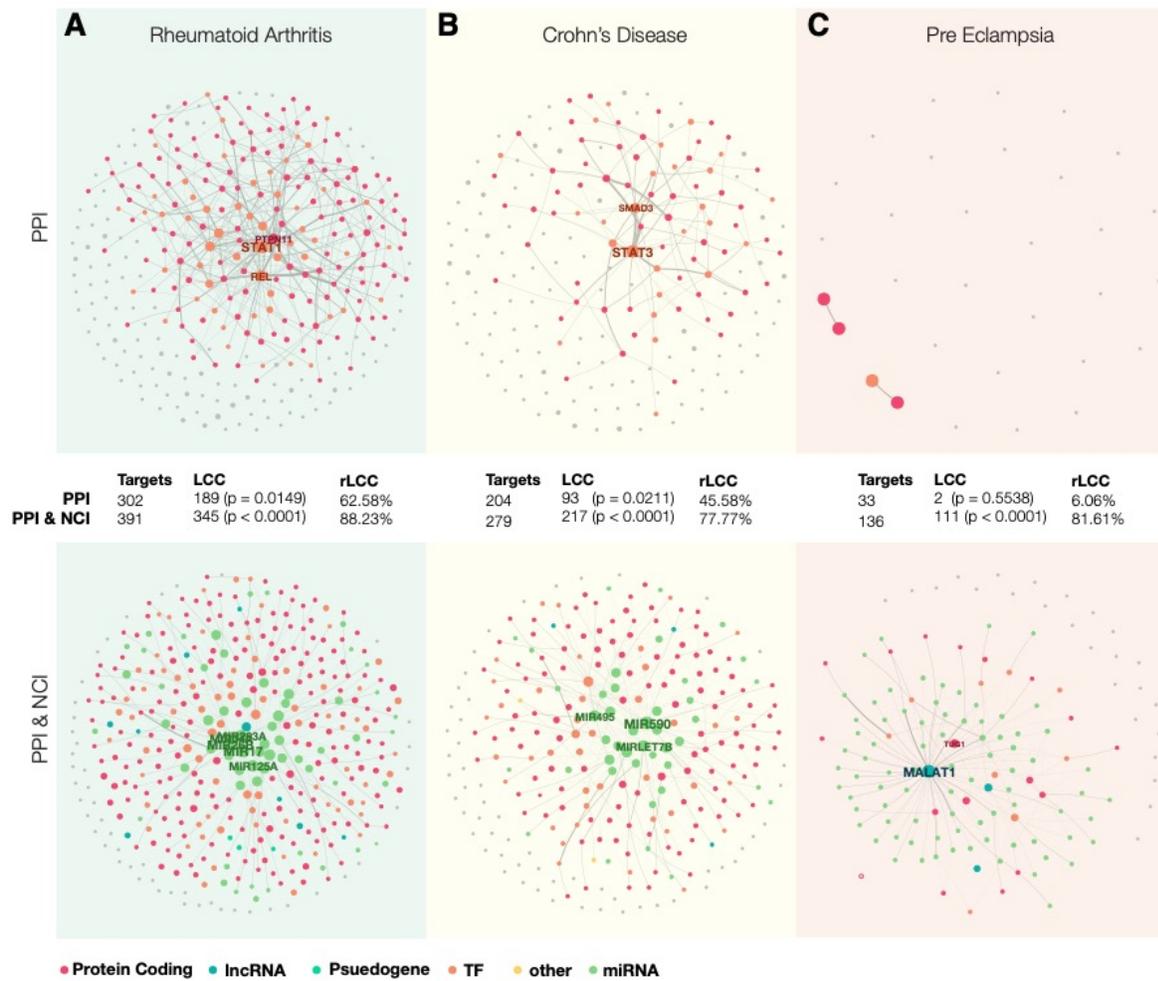

Figure 2 **Disease modules for Rheumatoid Arthritis (RA), Crohn's Disease (CD), and Pre-Eclampsia (PE). A)** Rheumatoid Arthritis Disease Module. In RA, both PPI and the PPI & NCI networks do form a significant LCC, however, the inclusion of ncRNAs into the network allows a much better retrieval of disease genes, increases the rLCC from rLCC$_{PPI}$: 62% to rLCC$_{PPI\ \&\ NCI}$: 88%. Gray nodes are not in the LCC of the complete interactome, while the colored nodes are present. **B)** Chron's Disease Disease Module. Both PPI and the PPI & NCI identify significant disease modules. However, by including ncRNAs into the disease module of CD, the rLCC increases from rLCC$_{PPI}$: 45% to rLCC$_{PPI\ \&\ NCI}$: 77%, increasing the disease gene retrieval. **C)** Pre-Eclampsia Disease Module. Pre-eclampsia shows that the inclusion of ncRNAs can change our ability to identify significant disease modules. In the PPI we are unable to define and identify a disease module, however, when in the PPI & NCI a disease module emerges (rLCC$_{PPI}$: 6%, and rLCC$_{PPI\ \&\ NCI}$: 81%).

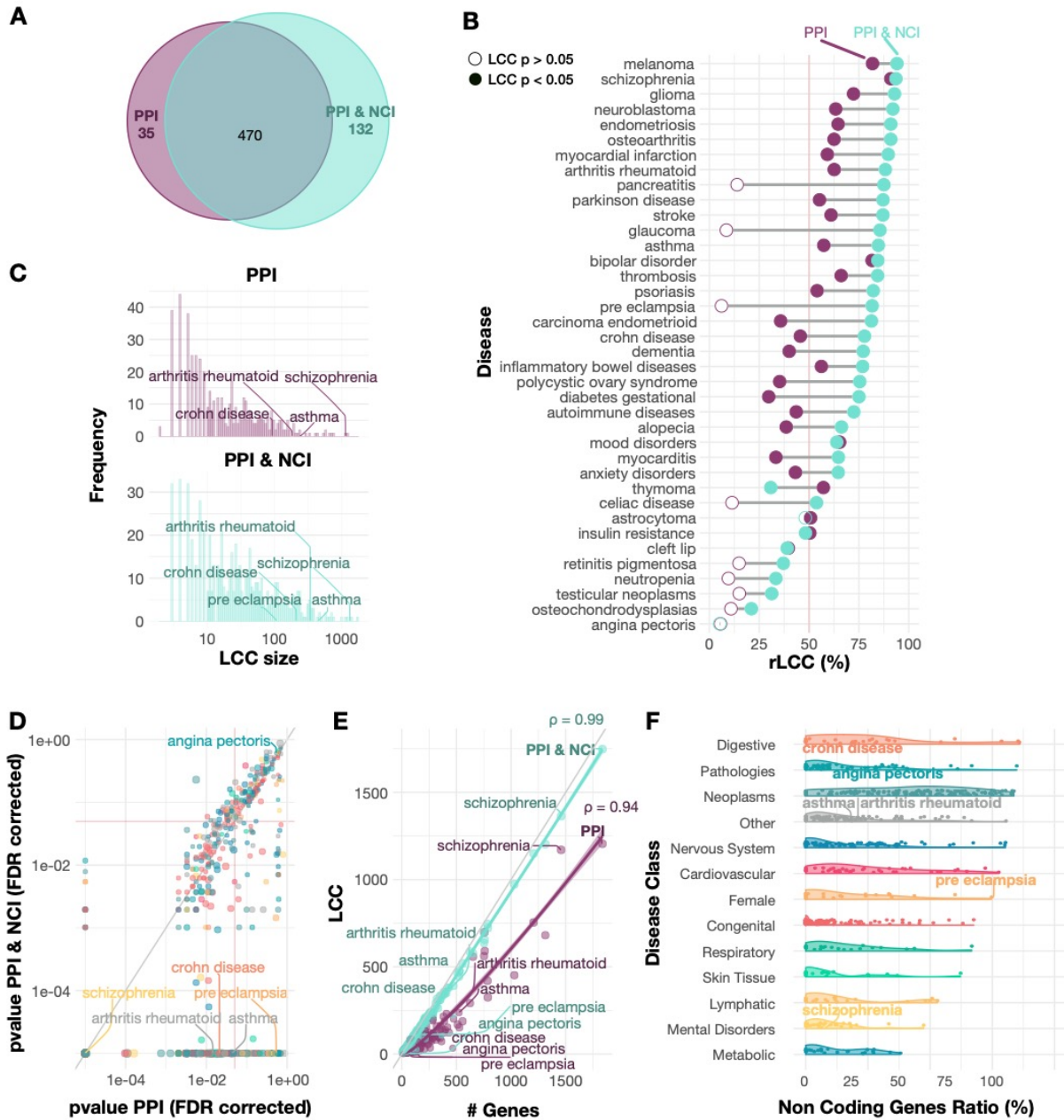

Figure 3 **Uncovering Gene-Disease Associations. A)** Overlap of Disease Modules Identification. Venn diagram representing the number of significant diseases in each network. In the PPI and the PPI & NCI, we can identify 470 disease modules, while solely the PPI identifies 35 disease modules (p < 0.05, FDR corrected). However, the PPI & NCI enables us to identify disease modules for 132 diseases that could not be identified previously. **B)** rLCC Change in the PPI and the PPI & NCI. Each line represents the rLCC size for a disease in the PPI and the PPI & NCI, along with disease-module significance. Empty circles represent diseases that do not have a significant LCC. The rLCC using the PPI & NCI is larger for most of the diseases than the rLCC for the PPI alone. In some cases, i.e, pre-eclampsia and glaucoma, we can identify disease modules that did not exist in the PPI. **C)** LCC Size Increases in the PPI & NCI. The histogram depicts LCC distribution for diseases in both PPI and PPI & NCI, the PPI (in purple) shows a distribution heavily shifted to the left, and the PPI & NCI (in turquoise) indicates a distribution that its values are shifting to the right, indicating that the average distribution of the LCCs in the combined network increases. **D)** The relationship between the p-value on the PPI and the PPI & NCI. the scatterplot shows the p-values in the PPI and the PPI & NCI for all diseases. On average, p-values of the disease module are smaller on the combined network, suggesting that the inclusion of NC elements improves the identification of the disease modules. The red lines indicate p-value = 0.05. **E)** Genes associated with disease and the LCC size. The scatterplot depicts the number of genes associated with a disease, and the disease-module size for both networks. The greater the number of known associated genes, the greater their LCC ($\rho_{PPI} = 0.94$; $\rho_{PPI \& NCI} = 0.99$, Pearson Correlation). We also find that PPI & NCI network has the LCCs closer to the total amount of genes associated with the disease, while the PPI has a smaller LCC compared to the number of genes per disease, suggesting an incompleteness of the disease module. **F)** Proportion of non-coding genes. The violin plot shows the percentage of ncRNAs associated with diseases classified in each disease category. We find that Digestive, Pathologies, and Neoplasms are the top three disease categories with the highest ratio of the non-coding genes, suggesting that ncRNAs play a role in the manifestation of those disease categories.

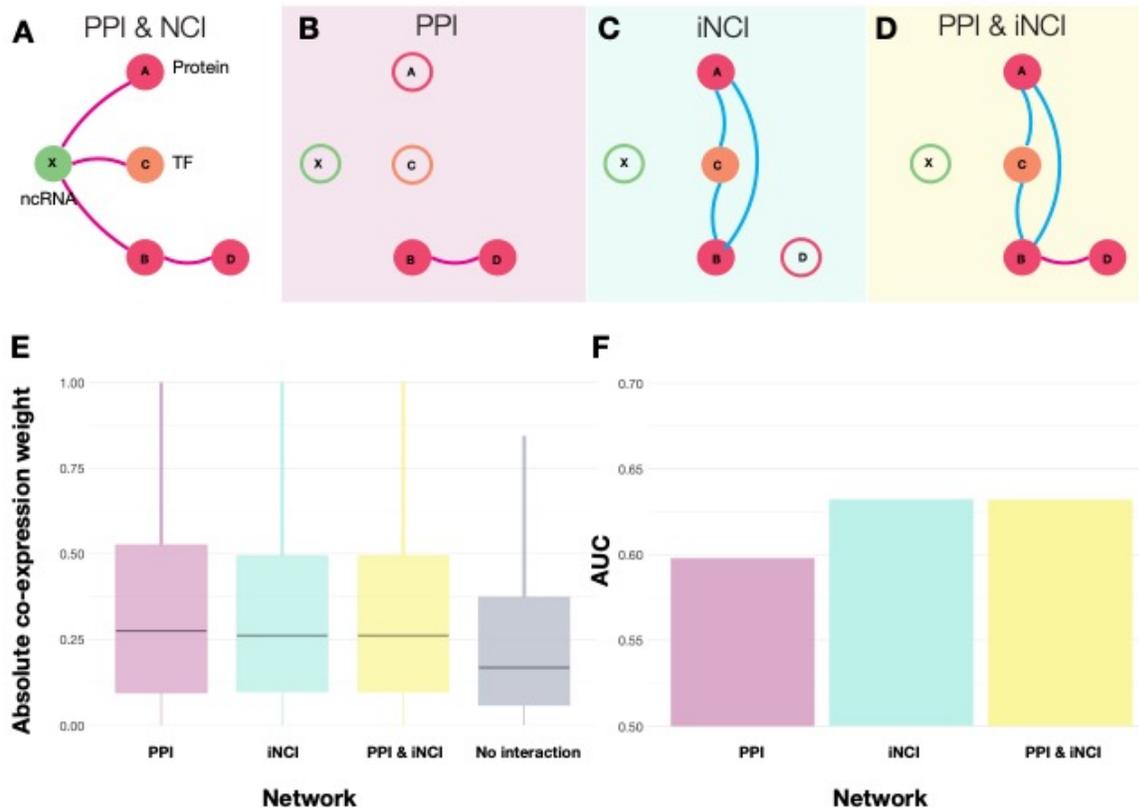

Figure 4 **Physically-Interacting genes are co-expressed. A)** Schematics of the Complete PPI and NCI Network. A non-coding RNA "X" interacts with three proteins "A", "B" and "C", and protein "B" binds to protein "D". **B)** The PPI Network Contain Only Interactions Among Proteins, and all interactions containing NC interactions are absent. **C)** The induced NCI (iNCI) contains indirect interactions. ncRNA "X" interacts with genes "A" and "B", therefore they are co-regulated by the same ncRNA, leading to an indirect interaction. In the same fashion, proteins "A" and "C" and "B" and "C" are also co-regulated by the same ncRNA "X", inducing a triangle among them. **D)** PPI & iNCI includes direct and indirect interactions. Combining the interactions identified in B and C. **E)** Genes with direct or indirect physical binding (PPI, PPI & NCI, or co-regulated by an ncRNA) have higher co-expression values than genes that do not physically interact. The boxplot indicates that the absolute Pearson correlation is higher when there is physical interaction, compared to then non-existing links, indicating an association between physical binding and strength of co-expression. **F)** Co-Expression Networks Can Predict Physical Interactions. We use the correlation values between two transcripts to predict a direct or indirect binding, finding that the inclusion of ncRNAs increases the AUC in the iNCI and the PPI & iNCI.

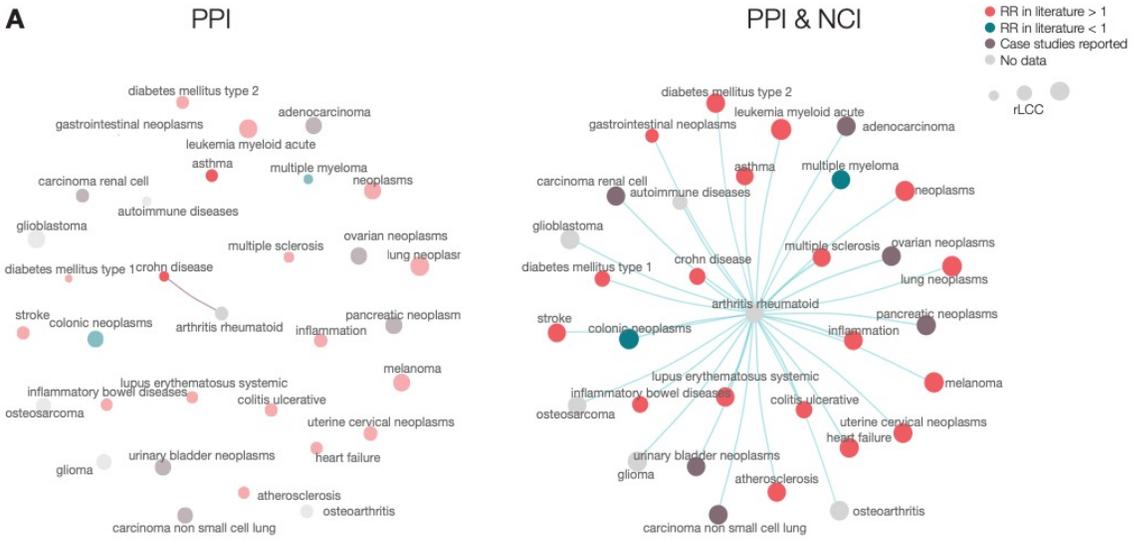
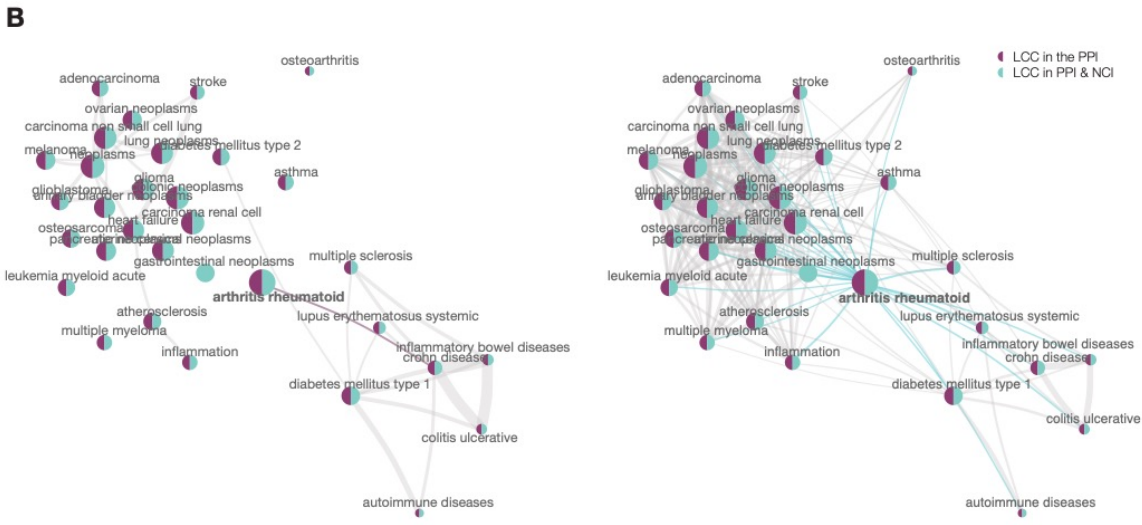
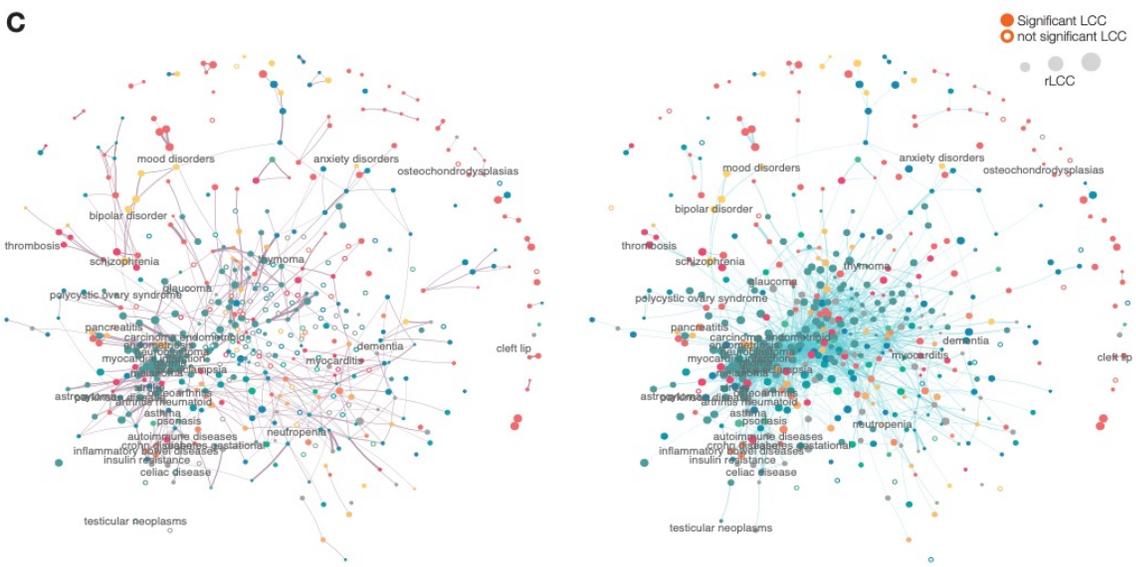

Figure 5 **Disease similarity on the two networks. A)** Rheumatoid Arthritis Disease Similarities And Possible Comorbidities. Two diseases are connected if they have a negative separation (meaning that they co-exist in the same network neighborhood) and if their overlap is significant. RA Disease-Disease network (LCC p < 0.05, $S_{a,b} < 0$, $S_{a,b}$ p < 0.05, Jaccard Index > 0.05, Hypergeometric Test p < 0.05; all p values are FDR corrected) is shown for both the PPI and the PPI & NCI. Diseases are colored according to literature references for comorbidity. We find in the PPI that only Chron's Disease is possible comorbidity, while the PPI & NCI predicts several other comorbid diseases, and most of them have a reported relative risk (RR) greater than 1, indicating comorbidity, few have RR smaller than 1, indicating a protective effect from those diseases. **B)** Rheumatoid Arthritis Comorbidities Are Also Connected. Diseases that have high comorbidity with RA are also connected. Similar to A, we find an expansion of disease associations in the PPI & NCI. The PPI alone identifies two distinct clusters: A neoplasm and an inflammation module. The inclusion of ncRNA into the PPI helps us identify how those two clusters (neoplasm and inflammation) are also interconnected with each other. **C)** Complete Map Of Disease-Disease Relationships**.** We show the complete disease-disease network, unveiling the comorbidity map between 466 diseases. Each disease with a significant module (full dots) has its node size representing the size of the disease module, and the link width is relative to the normalized absolute $S_{a,b}$ value. Note that the PPI & NCI network forms a connected component with all the 213 diseases that have a significant LCC, moreover, we see that neoplasm (represented in blue) are close and form a module. The PPI, in its turn, forms a connected component with only 100 diseases, and the combination of both gives us a component that includes 249 diseases.

2018. *Nucleic Acids Research* (2018). doi:10.1093/nar/gkx1037

# Supplemental Information

# Non-Coding RNAs Improve the Predictive Power of Network Medicine

Deisy Morselli Gysi and Albert-László Barabási

**Table of Contents**

# 1 NON-CODING RNAS: A SHORT REVIEW

The non-coding RNAs (ncRNAs) are responsible for different biological functions, such as the maintenance of gene expression. In brief, gene expression relies on DNA as a template for RNAs which then migrate to the cytoplasm, where they are translated into proteins. Yet, not all RNAs will be translated into proteins, and many ncRNAs are needed for diverse cell functions[1].

The **small nuclear RNAs (snRNAs)** are mainly involved with the mRNA splicing, Transfer RNAs (tRNAs) are the decoders from mRNAs into peptides or proteins, they recognize three nucleotide sequences in the mRNA (codons) and recruit amino acids in the same sequence to ribosomes; Ribosomal RNAs (rRNAs), the most abundant RNA molecules in the cell, are the building blocks for ribosomes, essential for protein translation (housekeeping RNAs). Some housekeeping RNAs can carry modifications, added by small nucleolar RNAs (snoRNAs)[2]. **Small double-stranded RNAs (dsRNAs)** are known to mediate post-transcriptional gene silencing of mRNAs, by a process known as RNA interference (RNAi)[3]. Additionally, some ncRNAs were also discovered to be involved in regulatory processes such as micro RNAs (miRNAs), piwi-associated RNAs (piRNAs), and small interfering RNAs (siRNAs).

**MicroRNAs (miRNAs)** are a family of small nonprotein coding RNAs, that contain ~ 22 nucleotides[4], and function as important regulators of gene expression (activating or repressing their translation[4,5], and mainly act at the post-transcriptional level[6] A mature miRNA can recognize their mRNA targets by base-pairing the seed region (2-8 of the miRNA nucleotides) to the complementary region on the targeted mRNA[7]. Each miRNA can have hundreds of targets and has been estimated that the human genome has more than 1000 miRNAs and about 10 to 30% of all human genes may be regulated by miRNAs[8]. One can use Argonaute (AGO) immunoprecipitation followed by high-throughput sequencing, termed AGO CLIP–seq, to examine the association between miRNA and mRNAs in a very specific manner[9,10]. However, this technique does not provide a direct linkage between miRNA and its mRNAs targets, but it uses as an intermediate the argonaute protein (AGO), which binds to miRNAs. To overcome this issue, a new technology called CLASH (crosslinking, ligation, and sequencing of hybrids) has been proposed[11], which ligates the 5′ end of an AGO-bound miRNA to the 3′ end of an AGO-bound mRNA, and as consequence, it provides direct linkage information of the two RNAs[12,13]. Because miRNAs function in a similar manner as Transcription Factors (TF), they have an intertwined and well-regulated network[14] that can contribute to disease development[15] such as asthma[16] and schizophrenia[17], when mutated or dysregulated miRNAs are highly associated with lack of function, such as neurogenesis[18].

**PIWI-interacting RNAs (piRNAs)**, were discovered from germlines, and are a little bit longer than miRNAs (24-32 nucleotides), piRNAs originate from single-strand RNAs, as opposed to miRNAs that originate from double-stranded RNAs and require post-processing. piRNAs act by directly cleaving mRNAs[19]. Until 2018 more than 8 million piRNAs have been discovered in humans[20]. Similarly, **small interfering RNA (siRNA)**, like miRNAs, are part of the RNAi class. Those are double-stranded ncRNAs with ~ 20-25 nucleotides and they are involved with the cell defense against undesired transcripts[21] and external invasion[22] and probably being our first biological mechanism that acted as the immune system[22]. siRNAs are highly sequence-specific, and in theory, can silence any disease-related genes, by mediating targeted mRNA. Given that, three drugs using siRNAs have been already approved to treat hereditary amyloidogenic transthyretin and acute hepatic porphyria diseases[22,23] and several others are in clinical trials for hemophilia, acute kidney injury, and others are being studied for IBD[24] and Ocular Diseases[25]; for a review see [26].

**Long non-coding RNAs (lncRNA)** are a family of nonprotein coding RNAs that exceed 200 nucleotides[27], present a poly-A tail, similar to mRNAs, and have potential to be spliced. Even though only a small number has been well-characterized[28], it is known that lncRNAs are involved in a wide range of biological functions, such as X-chromosome inactivation[29,30], imprinting[31,32], they can act as Gene Regulatory Factors (GRFs), and interact with one or more proteins[33–35], they also can bind to chromatin[36], enhancers[37] and act as sponges for miRNA[38]. Moreover, lncRNAs are associated with multiple diseases[39], including different types of cancers[40], autoimmune neuropathies[14], and neurodegenerative diseases[41]. Due to their function as GRF, lncRNAs can physically interact with proteins, and the measurement of the physical binding started by the low-throughput assays such as RNA electrophoretic mobility shift assay[42], RNA pull-down assay[43], oligonucleotide-targeted RNase H protection assay[44], and FISH co-localization[45]. However, these methods offer only limited information, mainly due to the many-to-many bindings that occur between proteins and RNAs. Nowadays, several different high-throughput methods exist, that can be classified into protein-focused and RNA-focused[28]. The protein-focused focus is on the binding of RNAs to a protein of interest and can be further classified into *in vitro* or *in vivo*. In the *in vitro* approaches, RNA libraries are tested against a protein, and high-affinity RNAs are isolated after rounds of selection. In the *in vivo* methods, RNAs bound to the protein of interest in a sample are pulled down using variants of immunoprecipitation techniques. In RNA-focused approaches, the goal is to identify all proteins bound to an RNA of interest. For a review with a complete description and comparison of the methods, see [28]. Of note, the lncRNA CRNDE has been identified as a promising target for the therapeutic treatment of prostate cancer by targeting miR-146a-5p[46].

## 2  DATABASE CONSTRUCTION

### 2.1  NON-CODING RNA DATABASES

We combine data from nine publicly available datasets that compile experimentally validated non-coding interactions. We describe each dataset, its data, and its normalization procedure. Table S 1 details their gene composition and interactions.

**DIANA Tools**[51] ranges from target prediction algorithms to databases of experimentally verified miRNA targets on coding and non-coding RNAs. In total it provides 91,249 interactions between 12,254 genes (10,087 proteins, 2,167 ncRNAs).

**lncBook**[49] carries information on lncRNAs, from functions, and associations to diseases and interactions from lncRNAs to other elements (proteins and miRNAs). The interactions are experimentally validated or predicted. In total, the database provides 21 interactions between 30 ncRNAs.

**lncRNome**[55] is a knowledge graph on human lncRNAs and provides predicted and experimentally validated interactions from lncRNAs and other RNAs. We only retrieved the database of experimentally validated targets, which offers information about 978 interactions between 521 genes (130 proteins, 391 ncRNAs).

**mirTARbase**[47,48] is a collection of experimentally validated miRNAs and their targets, validated by reporter assay, western blot, microarray, and next-generation sequencing experiments. After gene name conversion, it provides 6,667 interactions between 3,022 genes (2,511 proteins, 511 ncRNAs).

**miRecords**[56] is a manually curated database of experimentally validated miRNA-target interactions. It provides 975 validated interactions between 818 genes (651 proteins, 167 ncRNAs).

**miRNet**[53] aggregates information from miRTarBase v8.0, TarBase v8.0, and miRecords and allows for the selection of experimentally validated miRNA-targets. In total miRNet provides 2,546 interactions between 1,264 genes (917 proteins, 347 ncRNAs).

**NPinter4**[50] contains only experimentally validated interactions from ncRNA to DNA, TF, proteins, and other RNAs using CLIP-seq, AGO CLIP-seq, ChIRP-seq, and literature-mined interactions. For humans, it provides binding information for 1,153 interactions between 1,023 genes (538 proteins, 485 ncRNAs)

**RAIN**[54] contains experimentally validated and predicted interactions. Here we considered only experimentally validated interactions with a confidence score higher than 0.15 (as suggested by the authors), resulting in 391,209 interactions between 16,244 genes (13,271 proteins, 2,973 ncRNAs)

**RISE**[52] focuses on RNA-RNA interactions, which come from transcriptome-wide sequencing-based experiments such as PARIS, SPLASH, LIGRseq, and MARIO, and targeted studies like RIAseq, RAP-RNA, and CLASH. RISE also aggregates data from other databases with experimental validation, 64,084 validated interactions between 20,660 genes (15,559 proteins, 5,101 ncRNAs).

Note that, during the construction of the PPI some databases reported protein and ncRNA binds, we included those interactions only in the NCI. The Table S 1 provides a complete overview of how many genes and interactions exist in each final database, along with the number of experimentally validated interactions present in each network.

## 2.2 GENE DISEASE DATABASES

To link genes to diseases, we rely on several databases. We start by normalizing gene and disease names. A description of the gene-disease association in each database is presented in Table S 3.

**ClinGen**[57] is a worldwide effort to associate genes with diseases along with data curation from a panel of experts. Although it provides different levels of evidence, for each entry we focused only on the ones with strong or weak evidence. Overall, it accounts for 523 associations from 446 genes and 138 diseases.

**ClinVar**[58] reports relationships among human variations and phenotypes, with supporting evidence. It accounts for 6,769 associations between 3,847 genes and 917 diseases.

The **Comparative Toxicogenomics Database (CTD**[59]**)** is a well-established database of chemicals, genes, phenotypes, exposures, and diseases. All its 24,857 entries are annotated with publication information, allowing for transparency and for tracing any of its 7,327 genes and 7,711 diseases.

**Disease Enhancer**[60] is a manually curated database, based on literature, for disease-associated enhancers. It provides 518 associations between 303 genes and 121 diseases.

**DisGeNet**[61,62] integrates expert-curated databases with text-mined data and covers information on Mendelian and complex diseases. Here, we focus only on the curated databases based on genes or variants. The gene-based data source contains 36,317 associations, from 8,912 genes to 2,099 diseases.

**GWAS catalogue**[63] combines information across multiple GWAS studies extracted from literature and goes through a double curation process, where only associations with sufficient evidence are kept. We removed inferred and imputed SNPs and retrieved 8,957 associations, between 4,254 genes and 434 diseases.

**HMDD**[64]**, Human microRNA Disease Database**, focuses only on the association of miRNAs and diseases. Its data is experimentally validated through miRNA circulation, tissue differential expression, genetics, epigenetics, or targeted analysis. HMDD accounts for 13,662 associations between 916 miRNAs and 622 diseases.

**LncBook**[49]**,** used in the construction of the NCI, also provides information on lncRNAs' associations with diseases. It contains 1,478 associations between 669 genes and 246 diseases.

**lncRNADisease**[65,66] is a knowledge base that focuses on associations between diseases and lncRNA, it provides different levels of evidence for its associations. In our study, we focus on 3,146 associations with experimental evidence between 1,533 genes and 286 diseases.

**Leiden Open Variation Database (LOVD)**[67] is an integrative project connecting expert-curated variants and genes with multiple phenotypes. LOVD includes 2,931 associations from 2,113 genes to 664 diseases.

**Monarch** is an integrative project that connects phenotypes and genotypes in several species. It does provide experimental evidence for part of its results and includes text-mining information. In total, provides 19,542 associations between 8,059 genes and 1,058 diseases.

**Online Mendelian Inheritance in Man (OMIM**[68]**)** is a continuous catalog of human genes and traits, with a focus on the molecular relationship between genetic variation and phenotype. OMIM includes 3,517 associations based on peer-reviewed biomedical literature from 2,578 genes to 751 diseases.

**Orphanet** is a manually-curated database that focuses on experimentally validated associations from rare diseases to genes. It contains 3,787 associations from 2,555 genes to 721 rare diseases.

**PheGenI**[69] is the Phenotype-Genotype Integrator from NCBI and combines multiple of its hosted databases (such as Gene, dbGaP, OMIM, eQTL, and dbSNP). From this integrator, we retrieved 11,232 associations from 5,929 genes and 485 diseases.

**Psygenet**[70] is a database that focuses on psychiatric disorders. A gene is associated with a disease if it plays a role in the disease pathogenesis or is a marker for the disease. For that, the associations are identified using a text-mining tool, and later manually curated. It provides 2,924 associations between 1,324 genes in 40 mental disorders.

## 3 GENE CO-EXPRESSION NETWORKS

Gene co-expression networks are often used to shed light on the molecular mechanisms that underlie biological processes and how changes in those interactions can lead to a disorder[71]. These networks are inferred using the association between pairwise gene expression, measured using RNAseq or microarray. The association is often derived from a correlation, such as Pearson correlation, or a transformation, such as WGCNA[72] or Weighted Topological Overlap (wTO)[73,74], which calculates the co-expression between two genes based on their correlation normalized by their commonalities, and removes false positives. The former focuses on positive correlations, while the latter allows for the inclusion of positive and negative correlations. Here, we construct a gene co-expression network from whole blood samples provided by GTEx[75], using Pearson Correlation ($\rho$) and the wTO[73] ($\omega$).

Using the physical networks we derived from the PPI and the PPI & NCI (iNCI), we compared the co-expression values for both the Pearson Correlation and the wTO. We observed higher co-expression values on the physical networks (PPI, iNCI, and PPI & iNCI), leading us to ask whether the co-expression weights are predictive of physical binding. We calculated the Area Under the ROC (AUROC), for both co-expression values in the three physical networks, and we find that the AUCs increased from 0.59 in the PPI to 0.63 in the iNCI and PPI & iNCI networks for $\rho$ (Figure 4 F) and from 0.58 in the PPI to 0.63 in the iNCI and PPI & iNCI networks for $\omega$ (Figure S 6 B).

# 4 THE EFFECT ON NON-CODING RNAS IN DRUGS

Traditionally, drug-targets are proteins, currently, there is no ncRNA in DrugBank[76] listed as a drug-target. Yet, some ncRNAs are used as drugs – such as Bevasiranib[77], a small interfering RNA (siRNA) that targets the VEGF-A gene, currently in clinical trial for treating macular degeneration and diabetic retinopathy, or Inclisiran, an LDL cholesterol lowering siRNA that targets PCSK9, the first approved ncRNA-based drug [26,78].

Here, we explore the potential effect of ncRNAs on the drug-target network, by exploring the neighborhood of drug-targets. For that, we explore the differences in gene classes (protein-coding, TF, miRNA, lncRNA, etc) each drug-target interacts with, compared to non-drug-targets. We find that drug target proteins interact with more miRNAs, Protein-coding, and TF than proteins not targeted by drugs (Mann Whitney Test, p-adj < 0.05; FDR corrected, Table S 5), meaning that targets are, on median, more connected to miRNAs, protein-coding and TFs than not drugged proteins.

We begin by evaluating the effect of ncRNAs on drug proximity. For that, we calculate the average proximity of drug targets to the RA disease module in both networks. We first focus on three drugs used to treat RA: Adalimumab, Abatacept, and Penicillamine. Adalimumab targets the TNF gene which is also located in the RA disease module. In other words, both the PPI or the PPI & NCI disease modules already embed this gene in its network, meaning its average proximity is already zero. Abatacept, in turn, targets CD80 and CD86 in the PPI. The average proximity of those genes to the RA disease module is 2.34 in the PPI and 2.11 in the PPI & NCI, indicating that by adding non-coding interactions, we observe a reduction of the proximity values. Penicillamine targets only the gene SLCO1B1, having average proximity of two in the PPI, while the distance decreases to one in the PPI & NCI (Figure S 8 A). We next test the change in distance for all drugs indicated for RA in Drug Bank, finding that the average proximity reduces its proximity values from PPI to PPI & NCI. In other words, the inclusion of ncRNAs into the interactome reduces the distance between drug-targets and disease modules.

Even though the correlation between the proximity in the PPI and the PPI & NCI networks is high ($\rho = 0.97$), we still find a significant reduction of the average proximity in drugs indicated for RA the PPI & NCI ($p < 0.01$; Mann-Whitney's test). We next compare the average proximity between drugs indicated for RA compared to drugs without official indication for RA, finding that the proximity is statistically smaller for RA-indicated drugs than the ones not indicated to treat RA (Wilcoxon test; $p < 0.01$, (Figure S 8 B), both in the PPI and the PPI & NCI. Meaning that drugs used to treat RA are, on median, closer to the disease module than drugs not indicated for the RA treatment, in both PPI and PPI & NCI.

# 5 TABLES

Table S 1 **Descriptive of Databases Genes and Reported Binding Experimental Interactions.** Gene names were normalized to Gene Symbols and if no official name was found the gene was removed from the final database along with all its interactions. We kept only experimentally validated interactions and classified the interactions according to gene type (protein-coding - PC, non-coding – NC).

| DATABASE | GENES | | INTERACTIONS | | |
| --- | --- | --- | --- | --- | --- |
| | NC | PC | NC <-> NC | PC <-> NC | PC <-> PC |
| APID | 411 | 15930 | 32 | 3151 | 169730 |
| BIOGRID | 616 | 15308 | 31 | 9883 | 194221 |
| BIOPLEX | 386 | 13547 | 177 | 3912 | 104600 |
| COFRAC | 19 | 2959 | 3 | 205 | 13689 |
| DIANA | 2167 | 10087 | 15189 | 76060 | 0 |
| DIP | 2 | 1389 | 1 | 1 | 1285 |
| ENCODE | 1229 | 13808 | 1412 | 101569 | 23473 |
| HI-UNION | 165 | 8882 | 17 | 1703 | 61969 |
| HINT | 320 | 14576 | 66 | 2528 | 116789 |
| HIPPIE | 100 | 8520 | 17 | 1437 | 121691 |
| INATEDB | 153 | 5548 | 1 | 251 | 13378 |
| INSIDER | 17 | 3362 | 18 | 12 | 3939 |
| INSTRUCT | 206 | 11234 | 45 | 774 | 50456 |
| INTACT | 56 | 4567 | 10 | 131 | 9102 |
| INTERACTOME3D | 68 | 7327 | 108 | 135 | 15071 |
| INWEB | 67 | 10530 | 43 | 94 | 58507 |
| KINOMENETWORKX | 7 | 2340 | 0 | 25 | 7367 |
| LITBM17 | 20 | 6020 | 1 | 46 | 13320 |
| LNC_BOOK | 30 | 0 | 21 | 0 | 0 |
| LNCRNOME | 391 | 130 | 0 | 733 | 245 |
| MINT | 42 | 4871 | 6 | 121 | 10616 |
| MIRECORDS | 167 | 651 | 7 | 967 | 1 |
| MIRNET | 347 | 917 | 14 | 2532 | 0 |
| MIRTARBASE | 511 | 2511 | 51 | 6616 | 0 |
| NPINTERV4 | 485 | 538 | 210 | 860 | 83 |
| PHOSPHOSITEPLUS | 7 | 2840 | 0 | 7 | 7014 |
| PINA | 282 | 14895 | 12 | 958 | 163068 |
| QUBIC | 34 | 5271 | 7 | 241 | 27453 |
| RAIN | 2973 | 13271 | 7840 | 383213 | 156 |
| RISE | 5101 | 15559 | 2333 | 26660 | 35091 |

Table S 2 **Median and interquartile ranges for each gene category in all two networks.** TFs and protein-coding genes increase their degree median when we include the NCI in the PPI.

| GENE CATEGORY | PPI | PPI & NCI |
|---|---|---|
| **PROTEIN-CODING** | 30 [12; 64] | 54 [20; 107] |
| **TF** | 48 [20; 102] | 90 [41; 168] |
| **LNCRNA** |  | 3 [1; 6] |
| **MIRNA** |  | 52 [16; 224.75] |
| **OTHER** |  | 11 [3; 25] |
| **PSEUDOGENE** |  | 2 [1; 4] |

Table S 3 **Descriptive of Databases Genes and Reported Disease Associations.** Gene names were normalized to Gene Symbols and disease names were matched to MESH terms.

| | FINAL | | | ORIGINAL | | |
|---|---|---|---|---|---|---|
| **DATABASE** | Associations | Genes | Diseases | Associations | Genes | Diseases |
| **CLINGEN** | 523 | 446 | 138 | 896 | 695 | 352 |
| **CLINVAR** | 6,769 | 3,847 | 917 | 57,941 | 9,230 | 30,925 |
| **CTD** | 24,857 | 7,327 | 1,912 | 35,997 | 8,745 | 5,806 |
| **DISEASE ENHANCER** | 518 | 303 | 121 | 650 | 349 | 165 |
| **DISGENET (GDA)** | 36,317 | 8,912 | 2,099 | 95,959 | 10,147 | 11,181 |
| **GWAS CATALOG** | 8,957 | 4,254 | 434 | 70,192 | 12,786 | 2,339 |
| **HMDD** | 13,662 | 916 | 622 | 15,968 | 945 | 862 |
| **LNCBOOK** | 1,478 | 669 | 246 | 1,952 | 863 | 369 |
| **LNCRNA DISEASE** | 3,146 | 1,533 | 286 | 4,546 | 2,251 | 402 |
| **LOVD** | 2,931 | 2,113 | 664 | 6,611 | 3,910 | 4,860 |
| **MONARCH** | 19,542 | 8,059 | 1,058 | 33,549 | 11,531 | 8,593 |
| **OMIM** | 3,517 | 2,578 | 751 | 12,333 | 7,132 | 7,349 |
| **ORPHANET** | 3,787 | 2,555 | 721 | 8,252 | 4,324 | 3,784 |
| **PHEGENI** | 11,232 | 5,929 | 485 | 24,926 | 9,635 | 1,001 |
| **PSYGENET** | 2,924 | 1,354 | 40 | 3,621 | 1,493 | 108 |

Table S 4 **Genes with physical binding have higher co-expression in the physical networks.** Genes with a direct or indirect physical binding (PPI, PPI & NCI, or co-regulated by a ncRNA) have, in median, higher co-expression values than genes that do not physically interact. Kruskal-Wallis Test, Dunn's Post hoc test.

| | CORRELATION | | | WTO | | |
|---|---|---|---|---|---|---|
| **COMPARISON** | P | P adj | Kruskal-Wallis | P | P adj | Kruskal-Wallis |
| **DIRECT - NONE** | $p < 0.001$ | $p < 0.001$ |  | $p < 0.001$ | $p < 0.001$ |  |
| **INDIRECT - NONE** | $p < 0.001$ | $p < 0.001$ | $p < 0.001$ | $p < 0.001$ | $p < 0.001$ | $p < 0.001$ |
| **INDIRECT + DIRECT - NONE** | $p < 0.001$ | $p < 0.001$ |  | $p < 0.001$ | $p < 0.001$ |  |

Table S 5 **First Neighbors of drug-targets distribution.** Drug-Targets are more connected to miRNAs, TFs and Protein-coding Genes than non-drugged genes.

| | DRUGGED | NON-DRUGGED | P | P-ADJ |
|---|---|---|---|---|
| **PROTEIN-CODING** | 35 [15, 76] | 15 [4, 41] | <0.0001 | <0.0001 |
| **TF** | 9 [4, 21] | 6 [2, 16] | <0.0001 | <0.0001 |
| **LNCRNA** | 1 [1, 2] | 1 [1, 2] | 0.3875 | 0.3975 |
| **MIRNA** | 19 [5, 50] | 16 [4, 43] | <0.0001 | <0.0001 |
| **PSEUDOGENE** | 1 [1, 2] | 1 [1, 2] | 0.3237 | 0.2590 |

# 6 FIGURES

Figure S 1 **Network Validation: Gene-Interaction Overlap**. The upset plot depicts the number of binding interactions in each dataset, and how it overlaps across all PPI and NCI databases. Most of the interactions are database specific and do not occur in multiple sources, which provide us with complementary information. Interactions among proteins are represented in purple, while interactions involving at least one ncRNA are represented in turquoise.

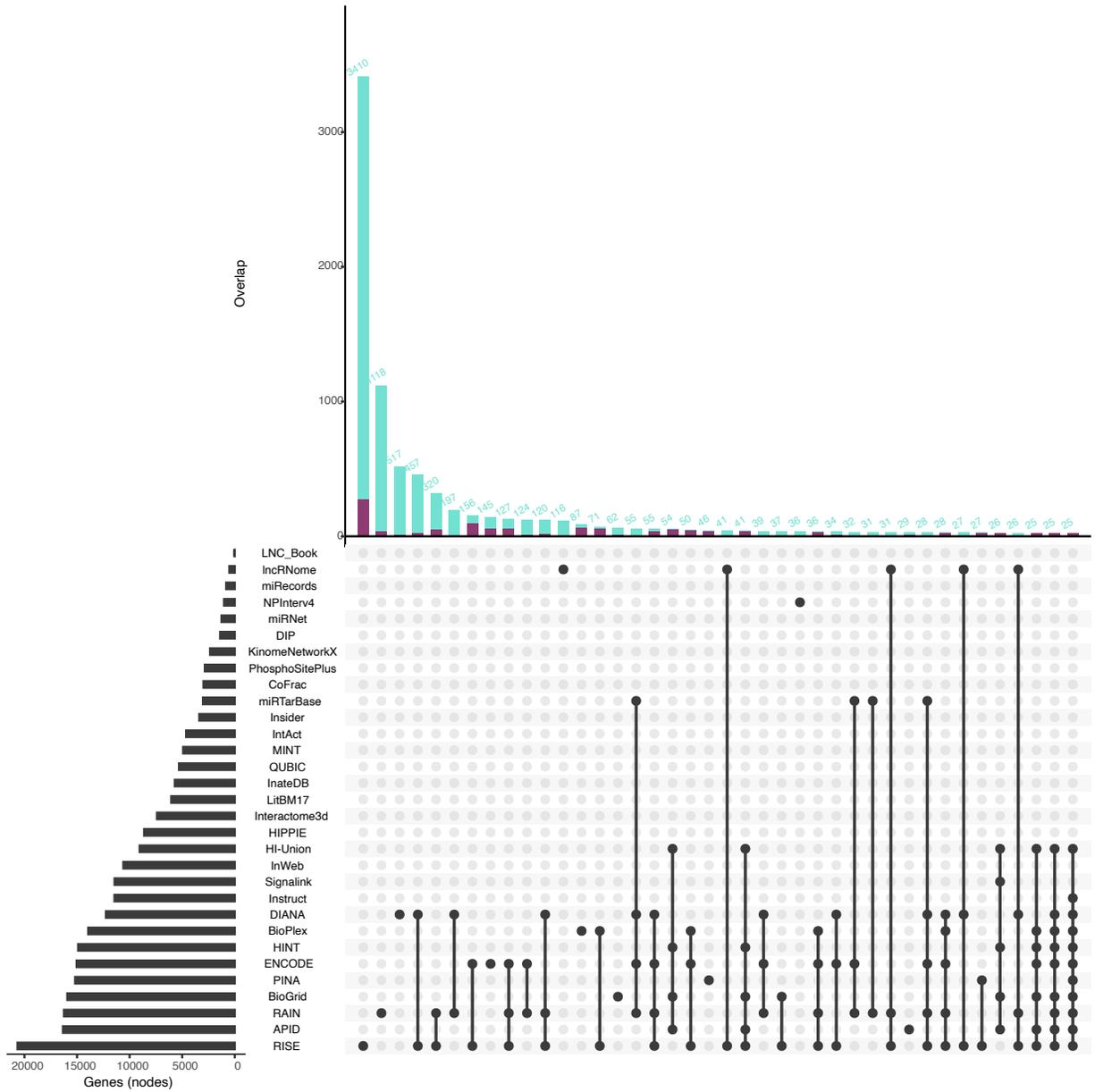

Figure S 2 **Network Validation: Gene Overlap**. The upset plot depicts the number of genes in each dataset, and how it overlaps across all PPI and NCI databases. Most of the genes are reported in multiple datasets. Proteins are represented in purple, while ncRNAs are represented in turquoise.

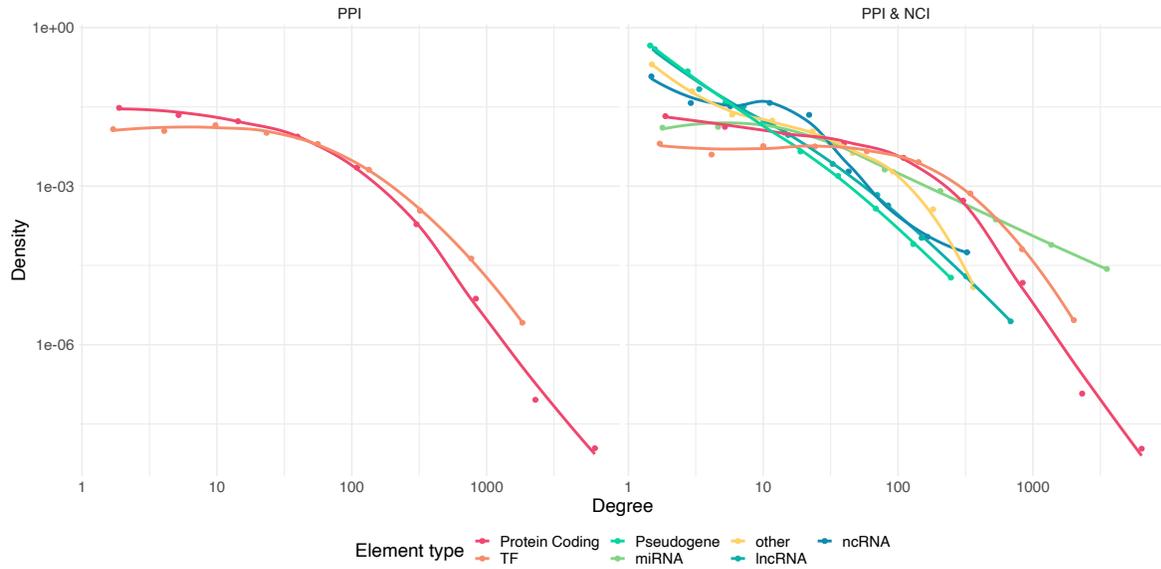

Figure S 3 **Degree distribution of different genomic elements.** The degree distribution on the three networks changes when we include non-coding elements into the PPI, the inclusion of miRNAs shows that they tend to have a higher degree, acting as regulatory factors and possibly as master regulators.

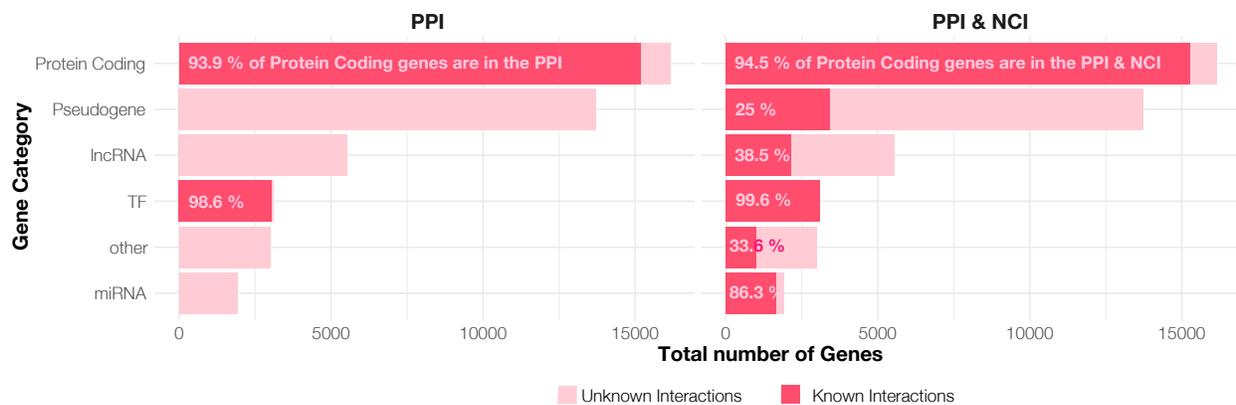

Figure S 4 **Gene completion mapped in the PPI and the combined network.** In the combined PPI & NCI we retrieve 86.3% of miRNAs, 99.6% of transcription factors and 38.5% of lncRNAs, increasing the interactions and coverage of the human transcriptomic.

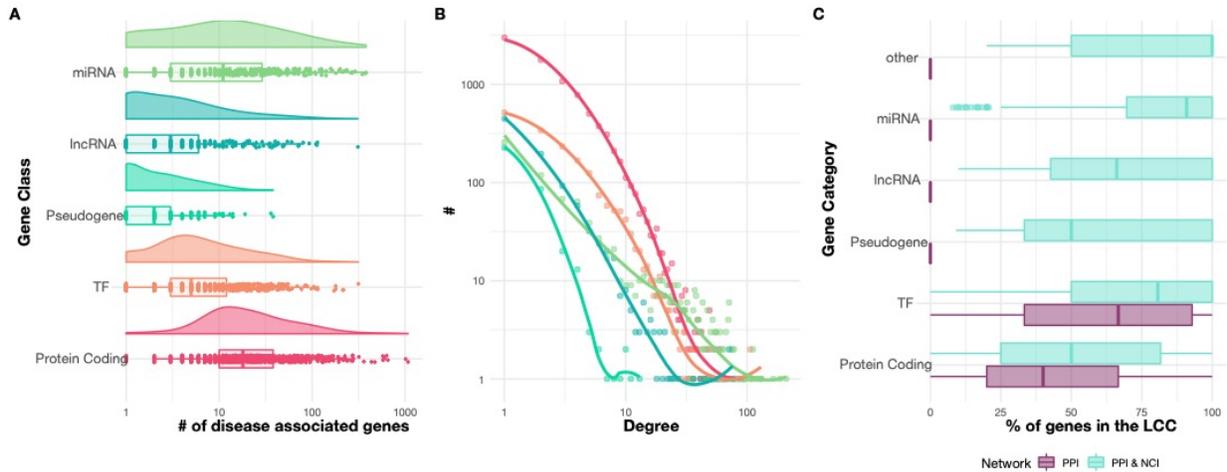

Figure S 5 **Disease Associated genes and their classification. A)** Rainplot of Number of disease-associated genes classified by gene category. We calculate the number of disease-associated genes in each disease, for each gene category. We find that Protein-Coding genes are the most associated with diseases, followed by miRNAs, TFs and lncRNAS. **B)** Degree distribution of gene-disease associations. miRNAs have a fat tail, indicating that few miRNAs can be associated with multiple diseases. **C)** The % of genes in each category found in the LCC. The inclusion of ncRNAs in the PPI allows us to increase the percentage of protein-coding genes retrieved in the disease modules from 40 to 50%. On median, 90% of miRNAs associated with a disease are found in the disease module.

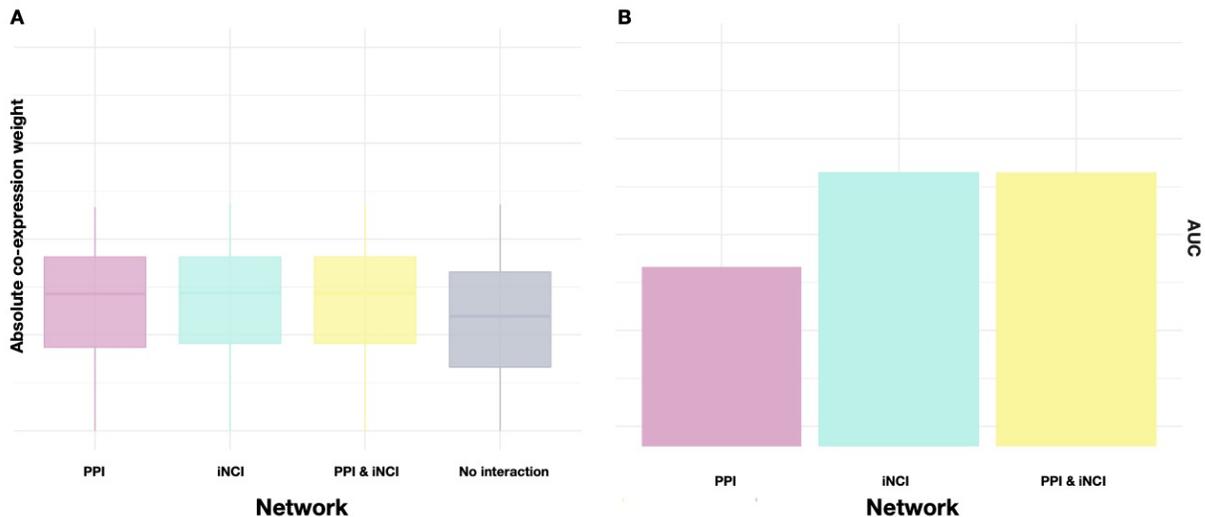

Figure S 6 **Absolute co-expression values (wTO) are higher in physical networks. A)** Genes with direct or indirect physical binding (PPI, PPI & NCI, or co-regulated by an ncRNA) have higher co-expression values than genes that do not physically interact in the wTO. The boxplot indicates that the absolute weighted Topological Overlap values are higher when there is physical interaction, compared to then non-existing links, indicating an association between physical binding and strength of co-expression. **B)** Co-Expression Networks Can Predict Physical Interactions. We use the wTO values between two transcripts to predict a direct or indirect binding, finding that the inclusion of ncRNAs increases the AUC in the iNCI and the PPI & iNCI.

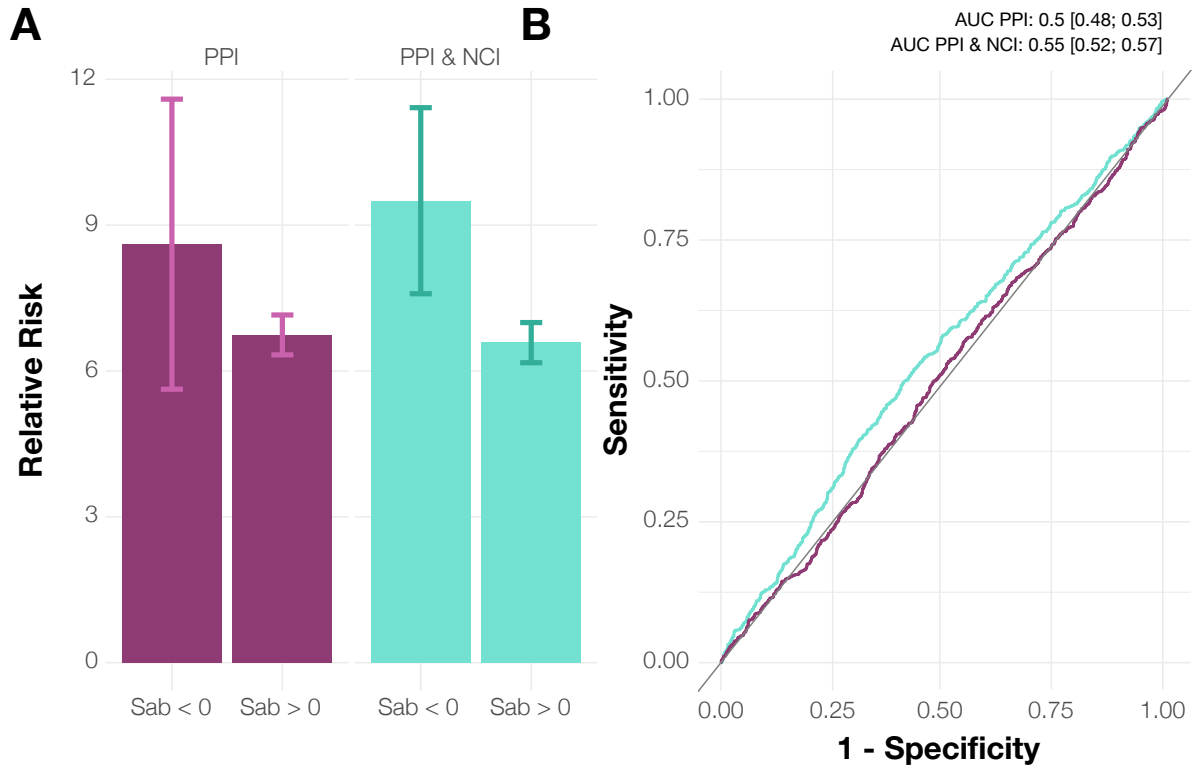

Figure S 7 **Disease Comorbidity and their separation. A)** Relative Risk is significatively higher for diseases closer to each other. Diseases with a negative $S_{ab}$ have significantly higher relative risks when compared to diseases with positive $S_{ab}$ for both the PPI and the PPI & NCI (Wilcoxon Test, $p < 0.05$). We observe for the PPI & NCI, on average, an increase on the RR of 9.5 for negative $S_{a,b}$ (se 2.93), and a decrease to 6.5 (se 0.41) for positive $S_{a,b}$ (Wilcoxon Test, $p > 0.05$). **B)** Network Separation is predictive of comorbidity. We use the network separation as a predictor of a significant Relative Risk > 1, finding that the PPI alone is as good as random (AUC = 0.5), while the PPI & NCI increases the AUC to 0.55, indicating that ncRNAs might hold the key for improving disease comorbidity and progression identification.

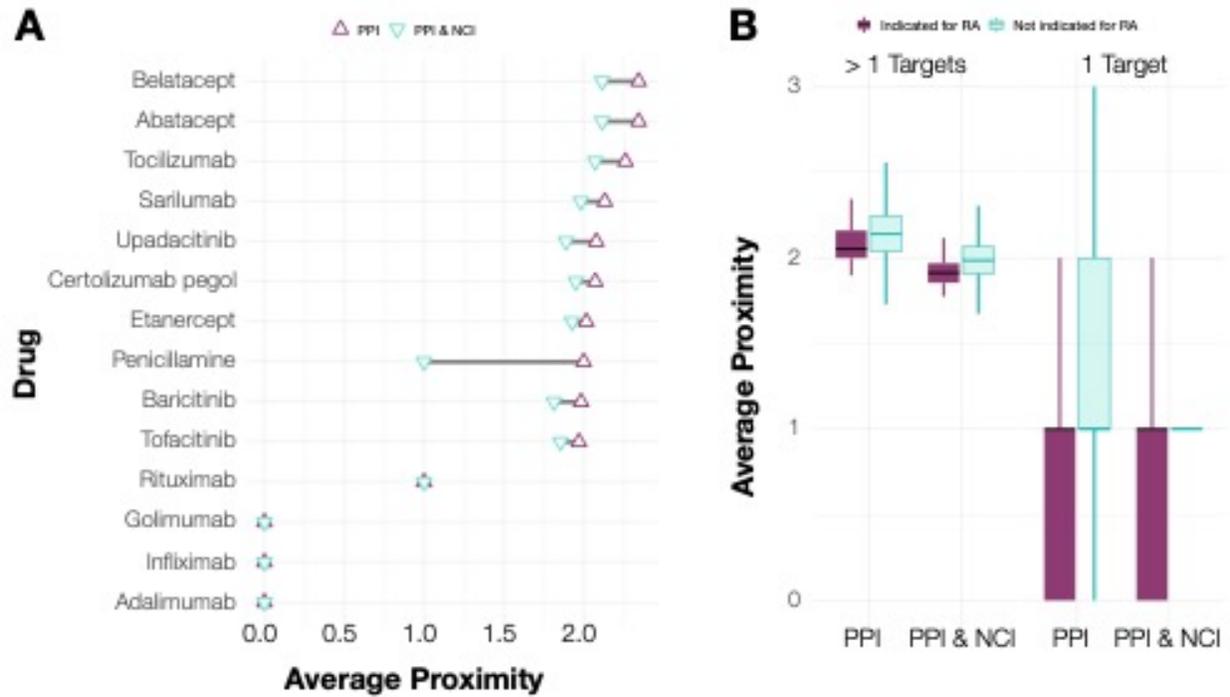

Figure S 8 **Average Proximity decreases for Rheumatoid Arthritis indicated drugs. A)** The dumbbell plot shows in the x-axis the average proximity for 14 drugs indicated to treat RA. In purple, the PPI values for each disease and in turquoise the PPI & NCI values. For all drugs with proximity higher than 1 we find that the PPI & NCI network decreases the distance of the drug-targets to the disease module. **B)** Average proximity for drug-targets with more than 1 target is statistically significant smaller for the PPI & NCI when compared to the PPI. The boxplots indicate a reduction on the proximity for drugs indicated to treat RA.